\documentclass[a4paper,12pt]{article}
\usepackage{amsmath,enumerate}
\usepackage{graphicx}

\newcommand{\lapprox}{%
\mathrel{%
\setbox0=\hbox{$<$}
\raise0.6ex\copy0\kern-\wd0
\lower0.65ex\hbox{$\sim$}
}}
\newcommand{\gapprox}{%
\mathrel{%
\setbox0=\hbox{$>$}
\raise0.6ex\copy0\kern-\wd0
\lower0.65ex\hbox{$\sim$}
}}

\begin{document}

\begin{center}

{\Large \bf Analyses of scalar potential and
lepton flavor violating decays in a model with $A_4$ symmetry}\\[20mm]

Raghavendra Srikanth Hundi\footnote{rshundi@phy.iith.ac.in} and
Itishree Sethi\footnote{ph15resch11004@iith.ac.in}\\
Department of Physics, Indian Institute of Technology Hyderabad,\\
Kandi - 502 284, India.\\[20mm]

\end{center}

\begin{abstract}

We have considered a model, originally proposed by Ma and Wegman,
where the mixing pattern in
neutrino sector is explained with three Higgs doublets, six Higgs triplets
and $A_4$ symmetry. The mixing pattern is explained with the help of
vacuum expectation values (VEVs) of the above mentioned doublets and triplets.
In order to study about the VEVs of the scalar fields,
we construct the full invariant scalar potential of this
model. After minimizing this scalar potential, we have found that two Higgs
triplets can acquire zero VEVs. In order to generate non-zero
VEVs to all the six Higgs triplets, we have added two more Higgs doublets
to the model. Thereafter we have demonstrated that the current neutrino
oscillation data can be consistently explained in our model. To study some
phenomenological implications of this model, we have worked out on the
branching ratios for lepton flavor violating decays.

\end{abstract}
\newpage

\section{Introduction}

Neutrino sector can give hints about physics beyond the standard model (SM)
\cite{bsm}. The masses of neutrinos are tiny as compared to other fermion
masses \cite{nu-rev}.
In order to explain the tiny masses for neutrinos, one has to extend the
SM. In addition to the masses of neutrinos, mixing pattern in neutrino
sector can also give a hint to physics beyond the SM. From the global fits
to neutrino oscillation data \cite{glo-fit}, the three neutrino mixing angles
are found
approximately close to the tri-bimaximal mixing (TBM) \cite{tbm}. To understand
this mixing pattern in neutrino sector, the SM should be extended with
additional symmetries and particle content \cite{discr}.

In this work, we consider the Ma-Wegman (MW) model
\cite{ma-weg}, where $A_4$ symmetry \cite{ma-raj} is introduced
to explain the neutrino mixing pattern. For early works on $A_4$ symmetry,
see Refs. \cite{a4mod}.
In the MW model, the scalar sector contains
three Higgs doublets and six Higgs triplets. Due to the presence of scalar Higgs
triplets, neutrinos acquire masses via type II seesaw mechanism \cite{t2se} in
this model. The above mentioned scalar fields and lepton doublets are charged
under $A_4$ symmetry in such a way that a realistic neutrino mixing pattern
can be explained. A unique feature of the MW model is that the six Higgs
triplets, which are responsible for obtaining the neutrino mixing pattern,
are charged under all possible irreducible representations of $A_4$ symmetry.
See Ref. \cite{ma-raj} for an introduction to $A_4$ symmetry. In Appendix A
we have summarized product rules among the irreducible representations
of $A_4$ symmetry.

The $A_4$ symmetry in the MW model is spontaneously broken
when the neutral component of doublet and triplet Higgs fields acquire
VEVs. The VEVs of the triplet Higgses generate
a mixing mass matrix
for neutrino fields. After diagonalizing this mass matrix, one can obtain
neutrino masses and mixing angles. In the work of MW model \cite{ma-weg}, this
diagonalization has been done after making some assumptions on the VEVs of
triplet Higgs fields, and thereby, it is concluded that neutrino masses
can have normal ordering. This problem of diagonalizing the neutrino
mass matrix of the MW model has been revisited in Ref. \cite{prwo}.
In the work of Ref. \cite{prwo}, after relaxing some of the assumptions made in
Ref. \cite{ma-weg} and also after using some approximation procedure
\cite{prwo2}, diagonalization
has been done for the neutrino mass matrix of the MW model. Thereafter,
it is concluded that both normal and inverted orderings for
neutrino masses are possible in the MW model, apart from
explaining the mixing pattern in neutrino sector.

As described above, the VEVs of scalar triplet
Higgses are responsible for generating the neutrino masses and mixing
angles in the MW model. One obtains the VEVs of scalar fields
after minimizing the invariant scalar potential among these fields. The
scalar potential in the MW model contains both the doublet
and triplet Higgs fields. Minimization for this scalar potential has not
been done before. On the other hand, minimization for the
invariant scalar potential containing only the Higgs doublets has been done in
Ref. \cite{ma-raj}, where it is shown that there exist a parameter region in
which the three Higgs doublets of this model acquire the same VEV. This is
known as the vacuum alignment of the Higgs doublets \cite{vacal}, which
is necessary to achieve in order to
diagonalize the charged lepton mass matrix, and thereby to explain the mixing
pattern in neutrino sector.

In this work, in order to see the implications of scalar potential on
neutrino masses and mixing pattern, we write the full invariant scalar
potential containing the three doublet and six triplet Higgses of the MW model.
After minimizing this scalar potential,
we have found that the two triplets, which are charged under the non-trivial
singlet representations of $A_4$ symmetry, acquire
zero VEVs. It is to remind here that in our previous work of Ref. \cite{prwo},
we assumed the VEVs of all triplet Higgses be non-zero and later showed that
neutrino oscillation data can be explained in the MW model. Moreover,
it is stated before that we followed a specific diagonalization
procedure in our previous work of Ref. \cite{prwo}. Now, in this work,
after finding that two Higgs triplets can acquire zero VEVs, with the
diagonalization procedure of Ref. \cite{prwo}, we have found that
the current neutrino oscillation data cannot be consistently explained.
To alleviate the above mentioned problem, we add two more Higgs doublets to
the MW model. After doing this, we show that at the minimum of the
scalar potential, all the six Higgs triplets can acquire non-zero VEVs.
As a result of this, we demonstrate that the neutrino oscillation data
can be fitted in this model for both normal and inverted neutrino mass
orderings. While doing the above mentioned minimization, we also address
the problem on vacuum alignment of the Higgs doublets. We show that sufficient
parameter region exist in this model, where the vacuum alignment of
the necessary Higgs doublets can be achieved.

After analyzing the scalar potential, it
is worth to study some phenomenological consequences of our model. We argue
below that the scalar fields of our model can drive lepton flavor violating
(LFV) processes
such as $\ell\to \ell^\prime\gamma$ and $\ell\to3\ell^\prime$.
Here, $\ell$ and $\ell^\prime$ are charged leptons
belonging to different families. None of the above mentioned LFV decays have
been observed in experiments, and as a result of that, upper limits on the
branching ratios of these processes have been obtained \cite{pdg}.
See Refs. \cite{relat}, for related studies on LFV processes in neutrino
mass models. In our model, the
above mentioned LFV decays are driven by the scalar fields which are
charged under the $A_4$ symmetry. Hence, one can expect that these decays
carry imprints of
$A_4$ symmetry. In this work, one of our interests is to study signatures
of $A_4$ symmetry in LFV decays. In a related direction to this, see
Ref. \cite{umaet}.

The scalar triplet Higgses
of our model drive LFV decays, since the Yukawa
couplings for lepton doublets are flavor violating in a type II seesaw
framework \cite{clp,lfv-rel}. We compute branching ratios for
the decays $\ell\to3\ell^\prime$ and $\ell\to\ell^\prime
\gamma$ in our model. The decays $\ell\to3\ell^\prime$ are driven by
doubly charged scalar triplets at tree level, whereas, the decays
$\ell\to\ell^\prime\gamma$  are driven by doubly and singly charged
scalar triplets at 1-loop level.
The above mentioned LFV decays can also be driven by scalar fields of doublet
Higgses, however, the contribution from these scalars has been neglected
in this work. We comment about this contribution later.
While computing the branching ratios for
the above mentioned decays, one needs to know the mass eigenstates of
the doubly and singly charged scalar triplets. These we obtain from the
invariant scalar potential of our model, which we have described above.
The branching ratios of the LFV decays in our work depend on Yukawa couplings
and the masses of above mentioned scalar fields. We have found that for
some decays the branching ratios are vanishingly small, if we assume
degenerate masses for triplet scalar fields. Another fact we have found is that,
due to the presence of $A_4$ symmetry,
some of the couplings between charged scalar triplets and leptons can
depend on one another. As a result of this, branching ratios for some
LFV decays can depend on each other. The above mentioned facts are some of the
signatures of $A_4$ symmetry in our model. Since the Yukawa couplings
depend on neutrino oscillation observables, numerically we study
the variation of these branching ratios in terms of neutrino mixing angles
and the $CP$ violating Dirac phase $\delta_{CP}$.

The paper is organized as follows. In the next section, we briefly
describe the MW model and present essential results from our
earlier work \cite{prwo} on this model. In Sec. 3, we construct the
full invariant scalar potential of this model and give our analysis
on the minimization of this potential. We study the implication of this
analysis on the neutrino mixing pattern by taking into account of the results
of our previous work \cite{prwo}. We demonstrate that by adding two
additional Higgs doublets, one can explain the neutrino mixing pattern
consistently in our model. In Sec. 4, we study the LFV decays of our model.
In Sec. 5, we describe future directions based on the phenomenology of our
model.
We conclude in the last section. In Appendix A, we have given the
product rules of $A_4$ symmetry, which are useful for making invariant
terms in our scalar potential. In Appendix B, we have listed all different
quartic terms of the scalar potential, which contain only the Higgs triplets.

\section{The MW model and essential results from it}

In this section, we describe the MW model \cite{ma-weg}. As stated
in the previous section, the method of diagonalizing the neutrino
mass matrix of this model has been improved in Ref. \cite{prwo}. The
essential results, related to neutrino masses and mixing angles, from the
work of Ref. \cite{prwo} are also presented in this section. These results
are used in our study on LFV decays, which is presented in Sec. 4.

The relevant fields of the MW model, along with their charge assignments
under the electroweak and $A_4$ symmetries are tabulated in Table 1.
\begin{table}[h]
\centering
\begin{tabular}{|c|c|c|c|c|c|c|c|c|c|} \hline
Field & $L_i=(\nu_{iL},\ell_{iL})^T$ & $\ell_{1R}$ & $\ell_{2R}$ & $\ell_{3R}$
& $\Phi_i$ & $\xi_1$ & $\xi_2$ & $\xi_3$ & $\xi_j$ \\ \hline
$A_4$ & $\underline{3}$ & $\underline{1}$ & $\underline{1}^\prime$ &
$\underline{1}^{\prime\prime}$ & $\underline{3}$ & $\underline{1}$ &
$\underline{1}^\prime$ & $\underline{1}^{\prime\prime}$ & $\underline{3}$
\\ \hline
$SU(2)_L$ & 2 & 1 & 1 & 1 & 2 & 3 & 3 & 3 & 3 \\ \hline
$U(1)_Y$ & $-\frac{1}{2}$ & $-1$ & $-1$ & $-1$ & $\frac{1}{2}$ & 1 & 1 & 1 & 1
\\ \hline
\end{tabular}
\caption{Fields in the lepton sector of the MW model \cite{ma-weg}.
Here, $i=1,2,3$ and $j=4,5,6$.}
\end{table}
With the charge assignments of Table 1, the
Yukawa couplings for charge leptons can be written as \cite{ma-raj}
\begin{equation}
{\cal L}=h_{ijk}\overline{L_i}\ell_{jR}\Phi_k+h.c.,\quad
\Phi_k=\left(\begin{array}{c} \phi_k^+\\\phi_k^0\end{array}\right).
\label{eq:yuklep}
\end{equation}
Here, $i,j,k=1,2,3$. $h_{ijk}$ are Yukawa couplings, whose form is determined
by $A_4$ symmetry, which can be seen in Ref. \cite{ma-raj}. Assuming that
the three Higgs doublets acquire the same VEV, after the electroweak symmetry
breaking, we get a mixing mass matrix for charged leptons. This mass matrix
can be diagonalized with the following transformations on the charged lepton
fields \cite{ma-raj}.
\begin{eqnarray}
&& \Psi_L\to U_L\Psi_L,\quad \Psi_R\to U_R\Psi_R,
\nonumber \\
&& \Psi_L=(\ell_{1L},\ell_{2L},\ell_{3L})^{T},\quad
\Psi_R=(\ell_{1R},\ell_{2R},\ell_{3R})^{T},
\nonumber \\
&& U_L = U_{CW} = \frac{1}{\sqrt{3}}\left(\begin{array}{ccc}
1 & 1 & 1 \\
1 & \omega & \omega^2 \\
1 & \omega^2 & \omega \end{array}\right),\quad
U_R = \left(\begin{array}{ccc}
1 & 0 & 0\\
0 & 1 & 0\\
0 & 0 & 1 \end{array}\right).
\label{eq:translep}
\end{eqnarray}
Here, $\omega=e^{2\pi i/3}$.

In the neutrino sector, the invariant Lagrangian is
\begin{eqnarray}
{\cal L} &=& y_1(\overline{L_1^c}i\sigma_2\xi_1L_1+
\overline{L_2^c}i\sigma_2\xi_1L_2+\overline{L_3^c}i\sigma_2\xi_1L_3)
\nonumber \\
&& +y_2(\overline{L_1^c}i\sigma_2\xi_2L_1+
\omega\overline{L_2^c}i\sigma_2\xi_2L_2+\omega^2\overline{L_3^c}i\sigma_2\xi_2
L_3)
\nonumber \\
&&+y_3(\overline{L_1^c}i\sigma_2\xi_3L_1+
\omega^2\overline{L_2^c}i\sigma_2\xi_3L_2+\omega\overline{L_3^c}i\sigma_2\xi_3
L_3)
\nonumber \\
&& +y(\overline{L_1^c}i\sigma_2\xi_6L_2+
\overline{L_2^c}i\sigma_2\xi_4L_3+\overline{L_3^c}i\sigma_2\xi_5L_1)+h.c.,
\label{eq:lag}
\\
&& \sigma_2=\left(\begin{array}{cc} 0 & -i \\ i & 0 \end{array}\right),\quad
\xi_k=\left(\begin{array}{cc}
\frac{\xi^+_k}{\sqrt{2}} & \xi^{++}_k \\
-\xi^0_k & -\frac{\xi^+_k}{\sqrt{2}}
\end{array}\right),k=1,\cdots,6.
\end{eqnarray}
Here, $y_1,y_2,y_3,y$ are dimensionless Yukawa couplings. Also,
$L_i^c$, where $i=1,2,3$, are charge conjugate doublets of $L_i$. The above
invariant Lagrangian can be obtained from the product rules of $A_4$
symmetry, which are given in Appendix A. After giving
VEVs to neutral component of $\xi_k$, from Eq. (\ref{eq:lag}), we get
mixing mass matrix for neutrino fields, which is given below \cite{ma-weg}.
\begin{eqnarray}
&& {\cal L} = -\frac{1}{2}\overline{\Psi^c}_\nu{\cal M}_\nu\Psi_\nu + h.c.,
\quad
\Psi_\nu = (\nu_{1L},\nu_{2L},\nu_{3L})^{T},\quad
\nonumber \\
&&{\cal M}_\nu = \left(\begin{array}{ccc}
a+b+c & f & e\\
f & a+\omega b+\omega^2c & d\\
e & d & a+\omega^2b+\omega c \end{array}\right),
\label{eq:Mnu}
\\
&& a=2y_1v_1^\prime, \quad b=2y_2v_2^\prime, \quad
c=2y_3v_3^\prime, \quad d=yv_4^\prime, \quad
e=yv_5^\prime, \quad f=yv_6^\prime.
\label{eq:vev}
\end{eqnarray}
Here, $\langle\xi^0_i\rangle=v^\prime_i$, $i=1,\cdots,6$. As stated
in the previous section, the masses for neutrinos are very small. In order
to obtain small masses for neutrinos, using the above relations, we can take
either the Yukawa couplings or the VEVs of Higgs triplets to be small.
In this work, we choose the VEVs of Higgs triplets to be small so that the
Yukawa couplings can be ${\cal O}(1)$. With this choice, we can notice
that LFV decays in this model are unsuppressed, and as explained in
the previous section, study of LFV decays is another topic of interest
in this work.

The matrix
in Eq. (\ref{eq:Mnu}) can be diagonalized after assuming $b-c,e,f$ to be small
and also after applying the following transformation on the neutrino fields
\cite{prwo}.
\begin{eqnarray}
&&\Psi_\nu\to U_{CW}U_{TBM}U_\epsilon\Psi_\nu,
\nonumber \\
&&U_{TBM} = \left(\begin{array}{ccc}
\sqrt{2/3} & 1/\sqrt{3} & 0 \\
-1/\sqrt{6} & 1/\sqrt{3} & -1/\sqrt{2} \\
-1/\sqrt{6} & 1/\sqrt{3} & 1/\sqrt{2} \end{array}\right),\quad
U_\epsilon = \left(\begin{array}{ccc}
1 & \epsilon_{12} & \epsilon_{13} \\
-\epsilon_{12}^* & 1 & \epsilon_{23} \\
-\epsilon_{13}^* & -\epsilon_{23}^* & 1 \end{array}\right).
\label{eq:trans}
\end{eqnarray}
In the unitary matrix $U_\epsilon$ \cite{ap-sc,im}, the $\epsilon$ parameters
are complex and the real and imaginary parts of these are assumed
\cite{prwo} to be less than or of the order of $\sin\theta_{13}\sim 0.15$
\cite{glo-fit}, where $\theta_{13}$ is a neutrino mixing angle.
Here, one can notice that $U_\epsilon$ is unitary only up to first order in
$\epsilon$ parameters.
From the above equation, we can see that $U_\epsilon$ gives a perturbation
to $U_{TBM}$, which can produce deviation from TBM mixing pattern. There
are other ways to parametrize these perturbations. However, in this work
we stick to the above mentioned parametrization, which is suggested in
Refs. \cite{ap-sc,im}. Now, while
diagonalizing the neutrino mass matrix in Eq. (\ref{eq:Mnu}), terms which are
of the order of $\sin^2\theta_{13}\sim\frac{m_s^2}{m_a^2}\sim 10^{-2}$
\cite{glo-fit} have been neglected \cite{prwo}. Here, $m_s$ and $m_a$ are
the square-root of solar and atmospheric mass-square differences
among the neutrino fields, respectively.
The central values for these mass-square differences are given below
\cite{glo-fit}.
\begin{equation}
m_s^2=m_2^2-m_1^2=7.5\times10^{-5}~{\rm eV}^2,\quad
m_a^2=\left\{\begin{array}{c}
m_3^2-m_1^2=2.55\times10^{-3}~{\rm eV}^2~~{\rm (NO)}\\
m_1^2-m_3^2=2.45\times10^{-3}~{\rm eV}^2~~{\rm (IO)}
\end{array}\right..
\label{eq:msquare}
\end{equation}
Here, $m_{1,2,3}$ are neutrino mass eigenvalues and NO(IO) represents
normal(inverted) ordering. In order to fit the above mass-square
differences, the neutrino mass eigenvalues can be taken as follows.
\begin{eqnarray}
&& {\rm NO}:\quad m_1\lapprox m_s,\quad m_2=\sqrt{m_s^2+m_1^2},\quad
m_3=\sqrt{m_a^2+m_1^2}.
\nonumber \\
&& {\rm IO}:\quad m_3\lapprox m_s,\quad m_1=\sqrt{m_a^2+m_3^2},\quad
m_2=\sqrt{m_s^2+m_1^2}.
\label{eq:numval}
\end{eqnarray}

As described previously, terms of the order of or
higher than that of $\sin^2\theta_{13}\sim\frac{m_s^2}{m_a^2}$ are
neglected in the diagonalization of $M_\nu$ of Eq. (\ref{eq:Mnu})
\cite{prwo}. As a result of this,
the neutrino mass eigenvalues in terms of model parameters
have been found to be \cite{prwo}
\begin{equation}
m_1=a+d-\frac{b+c}{2},\quad m_2=a+b+c,\quad m_3=-a+d+\frac{b+c}{2}.
\label{eq:rel1}
\end{equation}
The above expressions are valid in NO and IO cases. The relations for other
model parameters containing in $M_\nu$ depend on the neutrino mass ordering.
Expressions for these are given below \cite{prwo}.
\begin{eqnarray}
&& {\rm NO}:\quad e+f=0,\quad \frac{\sqrt{3}}{2}(b-c)=m_3\epsilon^*_{13},
\quad \frac{i}{\sqrt{2}}(e-f)=m_3\epsilon^*_{23}.
\nonumber \\
&& {\rm IO}:\quad \frac{e+f}{\sqrt{2}}=-m_1\epsilon_{12}+m_2\epsilon^*_{12},
\quad \frac{\sqrt{3}}{2}(b-c)=-m_1\epsilon_{13},\quad
\frac{i}{\sqrt{2}}(e-f)=-m_2\epsilon_{23}.
\nonumber \\
\label{eq:rel2}
\end{eqnarray}
Now, after diagonalizing the mass matrix $M_\nu$, one can get expressions
for neutrino mixing angles. The procedure for this is explained below.
After comparing the
transformations for charged leptons and neutrinos of Eqs. (\ref{eq:translep})
and (\ref{eq:trans}), the Pontecorvo-Maki-Nakagawa-Sakata (PMNS) matrix
can be written as
\begin{equation}
U_{PMNS} = U_{TBM}U_\epsilon.
\label{eq:pmns}
\end{equation}
The PMNS matrix is parametrized in terms of three neutrino mixing angles
and $\delta_{CP}$, in accordance with the PDG convention \cite{pdg}. Using this
parametrization in Eq.
(\ref{eq:pmns}) and after solving the relations of this matrix equation, the
leading order expressions for the three neutrino mixing angles and $\delta_{CP}$
are found to be \cite{prwo}
\begin{eqnarray}
&& \sin\theta_{12}=\frac{1}{\sqrt{3}}+\sqrt{\frac{2}{3}}Re(\epsilon_{12}),\quad
Im(\epsilon_{12})=0,
\nonumber \\
&& \sin\theta_{23}=-\frac{1}{\sqrt{2}}-\frac{1}{\sqrt{6}}Re(\epsilon_{13})+
\frac{1}{\sqrt{3}}Re(\epsilon_{23}), \quad Im(\epsilon_{13})=\sqrt{2}
Im(\epsilon_{23}),
\nonumber \\
&& \sin\theta_{13}=\left(\sqrt{\frac{2}{3}}Re(\epsilon_{13})+\frac{1}{\sqrt{3}}
Re(\epsilon_{23})\right)\cos\delta_{\rm CP} -
\left(\sqrt{\frac{2}{3}}Im(\epsilon_{13})+\frac{1}{\sqrt{3}}
Im(\epsilon_{23})\right)\sin\delta_{\rm CP},
\nonumber \\
&& \left(\sqrt{\frac{2}{3}}Re(\epsilon_{13})+\frac{1}{\sqrt{3}}
Re(\epsilon_{23})\right)\sin\delta_{\rm CP} +
\left(\sqrt{\frac{2}{3}}Im(\epsilon_{13})+\frac{1}{\sqrt{3}}
Im(\epsilon_{23})\right)\cos\delta_{\rm CP} = 0.
\nonumber \\
\label{eq:eps}
\end{eqnarray}
Here, $Re(\epsilon_{ij})$ and $Im(\epsilon_{ij})$ are real and imaginary
parts of $\epsilon_{ij}$, $i,j=1,\cdots,3$.

From Eq. (\ref{eq:eps}), we can notice that the imaginary part of
$\epsilon_{12}$
is zero. Using this in the case of IO, from Eq. (\ref{eq:rel2}), we get
$e+f\sim m_1\frac{m_s^2}{m_a^2}Re(\epsilon_{12})$. As described previously,
in the approximation procedure of Ref. \cite{prwo}, terms higher than the
order of
$\sin^2\theta_{13}\sim\frac{m_s^2}{m_a^2}$ are neglected. Hence, to the
leading order, in both NO and IO we get $e+f=0$. This implies
$v_5^\prime=-v_6^\prime$, which follows from Eq. (\ref{eq:Mnu}). Now, from Eq.
(\ref{eq:eps}), we can see that all $\epsilon$ parameters can be determined
in terms of three neutrino mixing angles and $\delta_{CP}$. Using this
fact and from Eqs.
(\ref{eq:rel1}) and (\ref{eq:rel2}), we can notice that all model
parameter of $M_\nu$ are determined in terms of neutrino oscillation
observables. Among these model parameters, except for $a$ and $d$, rest of
them depend on the neutrino mass ordering. Expressions for these parameters
are given below.
\begin{eqnarray}
&&{\rm NO~\& ~IO}:\quad a=\frac{m_1+m_2-m_3}{3},\quad d=\frac{m_1+m_3}{2}.
\nonumber \\
&&{\rm NO:}\quad b=\frac{m_2}{3}-\frac{m_1-m_3}{6}+
\frac{m_3\epsilon_{13}^*}{\sqrt{3}},
\quad c=\frac{m_2}{3}-\frac{m_1-m_3}{6}-\frac{m_3\epsilon_{13}^*}{\sqrt{3}},
\quad e=-\frac{im_3\epsilon_{23}^*}{\sqrt{2}}.
\nonumber \\
&&{\rm IO:}\quad b=\frac{m_2}{3}-\frac{m_1-m_3}{6}-
\frac{m_1\epsilon_{13}}{\sqrt{3}},
\quad c=\frac{m_2}{3}-\frac{m_1-m_3}{6}+\frac{m_1\epsilon_{13}}{\sqrt{3}},
\quad e=\frac{im_2\epsilon_{23}}{\sqrt{2}}.
\nonumber \\
\label{yuexp}
\end{eqnarray}
Using the above expressions in Eq. (\ref{eq:Mnu}), we can see that all
Yukawa couplings of the MW model can be determined in terms of neutrino
oscillation observables and the VEVs of Higgs triplets. Here, one can
notice that the coupling $y$ can be obtained from either $d$ or $e$. The
fact is that $v_4^\prime$ and $v_5^\prime$ are not independent parameters.
As a result of this, we can consider the following two cases in order
to determine $y$.
\begin{equation}
{\rm case~I:}\quad y=\frac{d}{v_4^\prime},\quad
{\rm case~II:}\quad y=\frac{e}{v_5^\prime}
\label{eq:case}
\end{equation}
In case I(II), $v_4^\prime(v_5^\prime)$ is independent parameter and
$v_5^\prime(v_4^\prime)$ is determined in terms of $v_4^\prime(v_5^\prime)$.
An interesting point is that if we choose $v_4^\prime$ as an independent
parameter, the coupling $y$ do not depend on neutrino mixing angles and
$\delta_{CP}$. On the other hand, in case II, $y$ depends on neutrino
mixing angles and $\delta_{CP}$. The above mentioned cases can make
a difference in the branching ratios for LFV decays of this model,
which is presented in Sec. 4.

\section{Analysis of scalar potential}

In the MW model \cite{ma-weg}, three Higgs doublets and six Higgs triplets
exist. From the previous section, we have seen that the VEVs of
Higgs triplets generate masses and mixing angles for neutrino fields.
The VEVs for these fields arise after minimizing the scalar potential
of this model. Hence, in this section, we write the full invariant scalar
potential of the MW model. Thereafter, we analyze the implications of this
potential on neutrino mixing.

\subsection{Scalar potential of the MW model}

The invariant scalar potential in the MW model can be written as
\begin{equation}
V_{MW}=V_0(\Phi)+V_1(\Phi,\xi)+V_Q(\xi).
\label{eq:vmw}
\end{equation}
Here, $V_0(\Phi)$ is a potential which depends only on the Higgs
doublets, whose form is already given in Ref. \cite{ma-raj}.
$V_1(\Phi,\xi)$ contains terms involving both
Higgs doublets and triplets. $V_Q(\xi)$ contains exclusively the quartic
interaction terms among the Higgs triplets. In the minimization of
the scalar potential, quartic terms in $V_Q(\xi)$ give negligibly small
corrections, due to the following reasons. From precision electroweak
tests \cite{pdg}, $\rho$ parameter gives a constraint on VEV of triplet
Higgs to be less than about 1 GeV. In the MW model, since three Higgs
doublets exist, we can choose the VEVs of Higgs doublets to be around
100 GeV. Hence, while doing the minimization, terms in $V_Q(\xi)$ are
at least suppressed by $10^{-4}$ as compared to that in $V_1(\Phi,\xi)$.
In our work, as stated in the previous section, we choose VEVs of Higgs
triplets to be much smaller than 1 GeV, say around 0.1 eV, in order to
explain the small neutrino masses. Shortly below, we give arguments for
making triplet Higgs VEVs to be as small as 0.1 eV.
For the above mentioned reasons,
we can notice that terms in $V_Q(\xi)$ give negligibly small contributions
to the minimization of the potential.
Hence, we omit these terms in our analysis. However, for
the sake of completeness, we present all the invariant terms of $V_Q(\xi)$
in Appendix B. In order to write the invariant terms of
$V_1(\Phi,\xi)+V_Q(\xi)$, we follow the work of Ref. \cite{maet}.
In Ref. \cite{maet}, invariant scalar potential under the electroweak
symmetry is given, for the case of one doublet and triplet Higgses.
To write the terms in $V_1(\Phi,\xi)+V_Q(\xi)$, we
generalize the potential given in Ref. \cite{maet},
by including three Higgs doublets, six Higgs triplets and $A_4$ symmetry.
In order to make the scalar potential invariant under $A_4$ symmetry,
we follow the product rules of $A_4$ symmetry, which are given in
Appendix A.

To write the scalar potential of the MW model,
we define the following quantities.
\begin{eqnarray}
&&(\Phi^\dagger\Phi)\equiv\Phi_1^\dagger\Phi_1+\Phi_2^\dagger\Phi_2
+\Phi_3^\dagger\Phi_3,\quad
(\Phi^\dagger\Phi)^\prime\equiv\Phi_1^\dagger\Phi_1+\omega^2\Phi_2^\dagger\Phi_2
+\omega\Phi_3^\dagger\Phi_3,
\nonumber \\
&&(\Phi^\dagger\Phi)^{\prime\prime}\equiv\Phi_1^\dagger\Phi_1
+\omega\Phi_2^\dagger\Phi_2+\omega^2\Phi_3^\dagger\Phi_3,\quad
(\xi^\dagger\xi)\equiv\xi_4^\dagger\xi_4+\xi_5^\dagger\xi_5
+\xi_6^\dagger\xi_6,
\nonumber \\
&&(\xi^\dagger\xi)^\prime\equiv\xi_4^\dagger\xi_4+\omega^2\xi_5^\dagger\xi_5
+\omega\xi_6^\dagger\xi_6,\quad
(\xi^\dagger\xi)^{\prime\prime}\equiv\xi_4^\dagger\xi_4
+\omega\xi_5^\dagger\xi_5+\omega^2\xi_6^\dagger\xi_6,
\nonumber \\
&&(\Phi^\dagger f(\xi)\Phi)\equiv\Phi_1^\dagger f(\xi)\Phi_1
+\Phi_2^\dagger f(\xi)\Phi_2+\Phi_3^\dagger f(\xi)\Phi_3,
\nonumber \\
&&(\Phi^\dagger f(\xi)\Phi)^\prime\equiv\Phi_1^\dagger f(\xi)\Phi_1
+\omega\Phi_2^\dagger f(\xi)\Phi_2+\omega^2\Phi_3^\dagger f(\xi)\Phi_3.
\end{eqnarray}
Here, $f(\xi)$ is a function depending on the Higgs triplet fields.
Now, we have \cite{ma-raj}
\begin{eqnarray}
V_0(\Phi)&=&m^2(\Phi^\dagger\Phi)+\frac{1}{2}\lambda_1(\Phi^\dagger\Phi)^2
+\lambda_2(\Phi^\dagger\Phi)^\prime(\Phi^\dagger\Phi)^{\prime\prime}
\nonumber \\
&&+\lambda_3\left[(\Phi_2^\dagger\Phi_3)(\Phi_3^\dagger\Phi_2)+
(\Phi_3^\dagger\Phi_1)(\Phi_1^\dagger\Phi_3)+(\Phi_1^\dagger\Phi_2)
(\Phi_2^\dagger\Phi_1)\right]
\nonumber \\
&&+\left\{\frac{1}{2}\lambda_4\left[(\Phi_2^\dagger\Phi_3)^2+
(\Phi_3^\dagger\Phi_1)^2+(\Phi_1^\dagger\Phi_2)^2\right]+h.c.\right\}.
\end{eqnarray}
In the above, $m^2$ has mass-square dimension and $\lambda$ parameters
are dimensionless.
The invariant terms in $V_1(\Phi,\xi)$ can be written as
\begin{eqnarray}
V_1(\Phi,\xi)&=&m_1^2~{\rm Tr}(\xi_1^\dagger\xi_1)
+m_2^2~{\rm Tr}(\xi_2^\dagger\xi_2)
+m_3^2~{\rm Tr}(\xi_3^\dagger\xi_3)+m_0^2~{\rm Tr}((\xi^\dagger\xi))
+\lambda_5^{(1)}(\Phi^\dagger\Phi){\rm Tr}(\xi_1^\dagger\xi_1)
\nonumber \\ &&
+\lambda_5^{(2)}(\Phi^\dagger\Phi){\rm Tr}(\xi_2^\dagger\xi_2)
+\lambda_5^{(3)}(\Phi^\dagger\Phi){\rm Tr}(\xi_3^\dagger\xi_3)
+\lambda_5^{(4)}(\Phi^\dagger\Phi){\rm Tr}((\xi^\dagger\xi))
\nonumber \\ &&
+\left\{\lambda_5^{(5)}(\Phi^\dagger\Phi)^\prime{\rm Tr}(\xi_1^\dagger\xi_3)
+\lambda_5^{(6)}(\Phi^\dagger\Phi)^\prime{\rm Tr}(\xi_2^\dagger\xi_1)
+\lambda_5^{(7)}(\Phi^\dagger\Phi)^\prime{\rm Tr}(\xi_3^\dagger\xi_2)
\right. \nonumber \\ && \left.
+\lambda_5^{(8)}(\Phi^\dagger\Phi)^\prime{\rm Tr}
((\xi^\dagger\xi)^{\prime\prime})
+\lambda_5^{(9)}\left[\Phi_2^\dagger\Phi_3{\rm Tr}(\xi_5^\dagger\xi_6)
+\Phi_3^\dagger\Phi_1{\rm Tr}(\xi_6^\dagger\xi_4)+
\Phi_1^\dagger\Phi_2{\rm Tr}(\xi_4^\dagger\xi_5)\right]
\right. \nonumber \\ && \left.
+\lambda_5^{(10)}\left[\Phi_2^\dagger\Phi_3{\rm Tr}(\xi_6^\dagger\xi_5)+
\Phi_3^\dagger\Phi_1{\rm Tr}(\xi_4^\dagger\xi_6)+
\Phi_1^\dagger\Phi_2{\rm Tr}(\xi_5^\dagger\xi_4)\right]
+h.c.\right\}
\nonumber \\ &&
+\lambda_6^{(1)}(\Phi^\dagger(\xi_1^\dagger\xi_1)\Phi)
+\lambda_6^{(2)}(\Phi^\dagger(\xi_2^\dagger\xi_2)\Phi)
+\lambda_6^{(3)}(\Phi^\dagger(\xi_3^\dagger\xi_3)\Phi)
+\lambda_6^{(4)}(\Phi^\dagger(\xi^\dagger\xi)\Phi)
\nonumber \\ &&
+\left\{\lambda_6^{(5)}(\Phi^\dagger(\xi_1^\dagger\xi_2)\Phi)^\prime
+\lambda_6^{(6)}(\Phi^\dagger(\xi_3^\dagger\xi_1)\Phi)^\prime
+\lambda_6^{(7)}(\Phi^\dagger(\xi_2^\dagger\xi_3)\Phi)^\prime
+\lambda_6^{(8)}(\Phi^\dagger(\xi^\dagger\xi)^\prime\Phi)^\prime
\right. \nonumber \\ && \left.
+\lambda_6^{(9)}\left(\Phi_1^\dagger\xi_6^\dagger\xi_4\Phi_3
+\Phi_2^\dagger\xi_4^\dagger\xi_5\Phi_1
+\Phi_3^\dagger\xi_5^\dagger\xi_6\Phi_2\right)
+\lambda_6^{(10)}\left(\Phi_1^\dagger\xi_4^\dagger\xi_5\Phi_2
+\Phi_2^\dagger\xi_5^\dagger\xi_6\Phi_3
\right.\right. \nonumber \\ && \left.\left.
+\Phi_3^\dagger\xi_6^\dagger\xi_4\Phi_1\right)
+\lambda_6^{(11)}\left(\Phi_1^\dagger\xi_6\xi_4^\dagger\Phi_3
+\Phi_2^\dagger\xi_4\xi_5^\dagger\Phi_1
+\Phi_3^\dagger\xi_5\xi_6^\dagger\Phi_2\right)
+\lambda_6^{(12)}\left(\Phi_1^\dagger\xi_4\xi_5^\dagger\Phi_2
\right.\right. \nonumber \\ && \left.\left.
+\Phi_2^\dagger\xi_5\xi_6^\dagger\Phi_3
+\Phi_3^\dagger\xi_6\xi_4^\dagger\Phi_1\right)
+\mu_1\left[\tilde{\Phi}_1^Ti\sigma_2\xi_1\tilde{\Phi}_1
+\tilde{\Phi}_2^Ti\sigma_2\xi_1\tilde{\Phi}_2
+\tilde{\Phi}_3^Ti\sigma_2\xi_1\tilde{\Phi}_3\right]
\right. \nonumber \\ && \left.
+\mu_2\left[\tilde{\Phi}_1^Ti\sigma_2\xi_2\tilde{\Phi}_1
+\omega\tilde{\Phi}_2^Ti\sigma_2\xi_2\tilde{\Phi}_2
+\omega^2\tilde{\Phi}_3^Ti\sigma_2\xi_2\tilde{\Phi}_3\right]
+\mu_3\left[\tilde{\Phi}_1^Ti\sigma_2\xi_3\tilde{\Phi}_1
\right.\right. \nonumber \\ && \left.\left.
+\omega^2\tilde{\Phi}_2^Ti\sigma_2\xi_3\tilde{\Phi}_2
+\omega\tilde{\Phi}_3^Ti\sigma_2\xi_3\tilde{\Phi}_3\right]
+\mu\left[\tilde{\Phi}_1^Ti\sigma_2\xi_5\tilde{\Phi}_3
+\tilde{\Phi}_2^Ti\sigma_2\xi_6\tilde{\Phi}_1
+\tilde{\Phi}_3^Ti\sigma_2\xi_4\tilde{\Phi}_2\right]
\right. \nonumber \\ && \left.
+h.c.\right\}.
\label{v1}
\end{eqnarray}
In the above equation, $m_{0,1,2,3}^2$ have mass-square dimensions,
$\mu$ parameters have mass dimensions and $\lambda$ parameters are
dimensionless. Here, $\tilde{\Phi}_k=i\sigma_2\Phi^*_k,k=1,2,3$.
We assume $\lambda$ parameters to be ${\cal O}(1)$. As stated before,
$\langle\phi_i^0\rangle\sim$ 100 GeV. $m_{0,1,2,3}^2\sim m_T^2$ give mass
scale for scalar triplet Higgses. Since we want these scalar triplet Higgses
to be produced in the LHC experiment, we take $m_{0,1,2,3}^2\sim m_T^2
\sim(100~{\rm GeV})^2$. Now,
after minimizing Eq. (\ref{v1}) with respect to triplet Higgses, naively
we expect the VEVs of these fields to be $\sim\mu_T\langle\phi_i^0\rangle^2
/m_T^2$. Here, $\mu_T$ represents any of the $\mu$ parameters of Eq.
(\ref{v1}). After using the above mentioned choices of the parameters,
we can see that the VEVs of triplet Higgses can be as small as 0.1 eV,
provided the $\mu$ parameters are suppressed to around 0.1 eV. By suppressing
the $\mu$ parameters, one can realize the hierarchy in the VEVs of doublet
and triplet Higgses. See Ref. \cite{clp}, for a loop induced mechanism
in order to explain the smallness of $\mu$ parameters.

The minimization of $V_0(\Phi)$ has been done in Ref. \cite{ma-raj} and
it is shown that $\langle\phi^0_i\rangle=v$ can be achieved for $i=1,2,3$.
In this work, doublet Higgses have interactions with triplet Higgses.
Since we are taking VEVs of triplet Higgses to be around 0.1 eV,
the contribution from $\langle V_1(\Phi,\xi)\rangle$ is negligibly
small in comparison to $\langle V_0(\Phi)\rangle$. Hence, in this work,
we get $\langle\phi^0_i\rangle\approx v$ for $i=1,2,3$. As stated previously,
this is known as the vacuum alignment of Higgs doublets, which is necessary
in order to diagonalize the charged lepton mass matrix, which is described
around Eq. (\ref{eq:translep}). Now, after minimizing the $V_1(\Phi,\xi)$
with respect to neutral components of $\xi_2$ and $\xi_3$, we get
\begin{equation}
(m_2^2+\lambda_5^{(2)}3|v|^2)\langle\xi_2^0\rangle=0,\quad
(m_3^2+\lambda_5^{(3)}3|v|^2)\langle\xi_3^0\rangle=0.
\label{eq:xi23}
\end{equation}
From the above equations, we get $v_2^\prime=v_3^\prime=0$. This implies
$b=c=0$. Using this in Eq. (\ref{eq:rel2}), we get $\epsilon_{13}=0$.
Thereafter, relations in Eq. (\ref{eq:eps}) can be solved for
$\delta_{CP}=\pi$, which is allowed for the case of NO by the current
neutrino oscillation data \cite{glo-fit}. As a result of this, at leading
order, we get the following constraint relation
\begin{equation}
\sin^2\theta_{23}=\frac{1}{2}+\sqrt{2}\sin\theta_{13}.
\end{equation}
The above constraint relation cannot be satisfied in the allowed
$3\sigma$ regions
for $\sin^2\theta_{23}$ and $\sin^2\theta_{13}$ \cite{glo-fit}.

The problem described in the previous paragraph arises due to the fact
that $\xi_2$ and $\xi_3$, which transform as $\underline{1}^\prime$ and
$\underline{1}^{\prime\prime}$ respectively under $A_4$, acquire zero VEVs.
On the other hand, the other Higgs triplets $\xi_1$ and $\xi_j$, 
which transform as $\underline{1}$ and $\underline{3}$ respectively under 
$A_4$, can acquire non-zero VEVs. Let us mention here that in Ref. \cite{im}
a model with $\xi_1$ and $\xi_j$ is presented in order to explain neutrino
mixing pattern. It is shown that the model of Ref. \cite{im} can consistently
explain neutrino mixing pattern and can predict normal ordering of masses
for neutrinos. Hence, one can see that the MW model, for the case of
$\langle\xi_{2,3}^0\rangle=0$, effectively reduces to that of Ref. \cite{im},
as far as neutrino mixing is concerned. As a result of this, even if
$\langle\xi_{2,3}^0\rangle=0$, the MW model can explain neutrino mixing
pattern but may only predict normal mass ordering for neutrinos. In this
regard, it is worth to see if the MW model can be modified in such a way that
it can explain both normal and inverted mass orderings for neutrinos.
In our earlier work \cite{prwo}, we had shown that the above mentioned
orderings are possible in the MW model, provided
$\langle\xi_{2,3}^0\rangle\neq0$. So to solve the above mentioned problem,
one needs to find a mechanism which can give $\langle\xi_{2,3}^0\rangle\neq0$
in the MW model.

One can notice that, because of the vacuum
alignment of Higgs doublets, the tri-linear couplings $\mu_2,\mu_3$ of
Eq. (\ref{v1}) do not contribute to $\langle\xi_2^0\rangle$ and
$\langle\xi_3^0\rangle$ after minimizing the scalar potential.
Whereas, the other tri-linear couplings $\mu_1,\mu$ can contribute to the VEVs
of rest of the Higgs triplets, even with the vacuum alignment of Higgs
doublets.
One cannot break the vacuum alignment of
Higgs doublets in the MW model, since it will affect the diagonalization of
charged lepton mass matrix, which in turn has an effect on the mixing pattern
in neutrino sector.
Hence, in order to give non-zero
VEVs to $\xi_{2,3}$, one can
introduce additional tri-linear couplings involving these fields.
We know that $\xi_{2,3}$ are charged under $\underline{1}^\prime,
\underline{1}^{\prime\prime}$ of $A_4$ symmetry.
With $\underline{1}^\prime,\underline{1}^{\prime\prime}$, the following are
the only two singlet combinations, which can represent tri-linear terms
in the potential: $\underline{1}^\prime\times\underline{1}^\prime
\times\underline{1}^\prime$, $\underline{1}^{\prime\prime}\times
\underline{1}^{\prime\prime}\times\underline{1}^{\prime\prime}$. Hence,
in the additional tri-linear couplings containing $\xi_{2,3}$, the Higgs
doublets should be charged under $\underline{1}^\prime$ and
$\underline{1}^{\prime\prime}$ of $A_4$ symmetry. As a result of this,
we propose additional Higgs doublets $\Phi_5$
and $\Phi_6$ which transform as $\underline{1}^\prime$ and
$\underline{1}^{\prime\prime}$ respectively under $A_4$ symmetry.
Now, one can see that the following terms can exist in the scalar potential,
which can give non-zero VEVs to $\xi_2$ and $\xi_3$:
$\tilde{\Phi}_6^{T}i\sigma_2\xi_2\tilde{\Phi}_6$,
$\tilde{\Phi}_5^{T}i\sigma_2\xi_3\tilde{\Phi}_5$. However,
the Higgs doublets $\Phi_5$ and $\Phi_6$ can give rise to extra terms in
the scalar potential with the fields $\Phi_i,i=1,2,3$. These extra
terms can affect the vacuum alignment of Higgs doublets of the MW model.
We study these topics in the next subsection.

\subsection{Extension of the MW model with two additional Higgs doublets}

As described previously, in order to get non-zero VEVs to $\xi_2$ and
$\xi_3$, we add the Higgs doublets $\Phi_5$ and $\Phi_6$ to the MW
model. Since these Higgs doublets are charged under $\underline{1}^\prime$
and $\underline{1}^{\prime\prime}$ of $A_4$ symmetry, with the charge
assignments given in Table 1, one can notice that they do not generate
Yukawa couplings for charged leptons and neutrinos.
However, the Higgs doublets $\Phi_{5,6}$
can have interactions with the other Higgs doublets
$\Phi_{1,2,3}$ and also with the Higgs triplets of this model.
As a result of this,
the scalar potential of the MW model, which is given in Eq. (\ref{eq:vmw}),
will change to
\begin{equation}
V=V_{MW}+V_0^\prime(\Phi)+V_1^\prime(\Phi,\xi).
\end{equation}
Here, $V_0^\prime(\Phi)$ and $V_1^\prime(\Phi,\xi)$ contain terms between
$\Phi_{5,6}$ and
already existing scalars of the MW model. Their forms are
given below.
\begin{eqnarray}
V_0^\prime(\Phi)&=&m_5^2\Phi_5^\dagger\Phi_5+m_6^2\Phi_6^\dagger\Phi_6
+\frac{1}{2}\lambda_1^{(1)}(\Phi_5^\dagger\Phi_5)^2
+\frac{1}{2}\lambda_1^{(2)}(\Phi_6^\dagger\Phi_6)^2
+\lambda_1^{(3)}(\Phi_5^\dagger\Phi_5)(\Phi_6^\dagger\Phi_6)
\nonumber \\ &&
+\lambda_1^{(4)}(\Phi^\dagger\Phi)(\Phi_5^\dagger\Phi_5)
+\lambda_1^{(5)}(\Phi^\dagger\Phi)(\Phi_6^\dagger\Phi_6)
+\lambda_2^{(1)}(\Phi_5^\dagger\Phi_6)(\Phi_6^\dagger\Phi_5)
\nonumber \\ &&
+\left\{\lambda_2^{(2)}(\Phi^\dagger\Phi)^\prime(\Phi_6^\dagger\Phi_5)
+h.c.\right\}
\nonumber \\ &&
+\lambda_3^{(1)}\left[(\Phi_5^\dagger\Phi_1)(\Phi_1^\dagger\Phi_5)
+(\Phi_5^\dagger\Phi_2)(\Phi_2^\dagger\Phi_5)
+(\Phi_5^\dagger\Phi_3)(\Phi_3^\dagger\Phi_5)\right]
\nonumber \\ &&
+\lambda_3^{(2)}\left[(\Phi_6^\dagger\Phi_1)(\Phi_1^\dagger\Phi_6)
+(\Phi_6^\dagger\Phi_2)(\Phi_2^\dagger\Phi_6)
+(\Phi_6^\dagger\Phi_3)(\Phi_3^\dagger\Phi_6)\right]
+\left\{\frac{1}{2}\lambda_4^{(1)}\left[(\Phi_5^\dagger\Phi_1)^2
\right.\right. \nonumber \\ && \left.\left.
+\omega^2(\Phi_5^\dagger\Phi_2)^2+\omega(\Phi_5^\dagger\Phi_3)^2\right]
+\frac{1}{2}\lambda_4^{(2)}\left[(\Phi_6^\dagger\Phi_1)^2
+\omega(\Phi_6^\dagger\Phi_2)^2+\omega^2(\Phi_6^\dagger\Phi_3)^2\right]
\right. \nonumber \\ && \left.
+\lambda_4^{(3)}\left[(\Phi_5^\dagger\Phi_1)(\Phi_6^\dagger\Phi_1)+
(\Phi_5^\dagger\Phi_2)(\Phi_6^\dagger\Phi_2)+
(\Phi_5^\dagger\Phi_3)(\Phi_6^\dagger\Phi_3)\right]
\right. \nonumber \\ && \left.
+\lambda_4^{(4)}\left[(\Phi_5^\dagger\Phi_1)(\Phi_1^\dagger\Phi_6)+
\omega(\Phi_5^\dagger\Phi_2)(\Phi_2^\dagger\Phi_6)+
\omega^2(\Phi_5^\dagger\Phi_3)(\Phi_3^\dagger\Phi_6)\right]+h.c.\right\}.
\end{eqnarray}
\begin{eqnarray}
V_1^\prime(\Phi,\xi)&=&\Phi_5^\dagger\Phi_5\left[
\lambda_5^{(11)}{\rm Tr}(\xi_1^\dagger\xi_1)
+\lambda_5^{(12)}{\rm Tr}(\xi_2^\dagger\xi_2)
+\lambda_5^{(13)}{\rm Tr}(\xi_3^\dagger\xi_3)
+\lambda_5^{(14)}{\rm Tr}((\xi^\dagger\xi))\right]
\nonumber \\ &&
+\Phi_6^\dagger\Phi_6\left[
\lambda_5^{(15)}{\rm Tr}(\xi_1^\dagger\xi_1)
+\lambda_5^{(16)}{\rm Tr}(\xi_2^\dagger\xi_2)
+\lambda_5^{(17)}{\rm Tr}(\xi_3^\dagger\xi_3)
+\lambda_5^{(18)}{\rm Tr}((\xi^\dagger\xi))\right]
\nonumber \\ &&
+\Phi_5^\dagger\Phi_6\left[
\lambda_5^{(19)}{\rm Tr}(\xi_1^\dagger\xi_3)
+\lambda_5^{(20)}{\rm Tr}(\xi_2^\dagger\xi_1)
+\lambda_5^{(21)}{\rm Tr}(\xi_3^\dagger\xi_2)
+\lambda_5^{(22)}{\rm Tr}((\xi^\dagger\xi)^{\prime\prime})\right]
\nonumber \\ &&
+\Phi_6^\dagger\left[\lambda_6^{(13)}\xi_1^\dagger\xi_2
+\lambda_6^{(14)}\xi_3^\dagger\xi_1+\lambda_6^{(15)}\xi_2^\dagger\xi_3
+\lambda_6^{(16)}(\xi^\dagger\xi)^\prime\right]\Phi_5
+\Phi_6^\dagger\left[\lambda_6^{(17)}\xi_2\xi_1^\dagger
\right. \nonumber \\ && \left.
+\lambda_6^{(18)}\xi_1\xi_3^\dagger+\lambda_6^{(19)}\xi_3\xi_2^\dagger
+\lambda_6^{(20)}(\xi\xi^\dagger)^\prime\right]\Phi_5
+\mu_4\tilde{\Phi}_6^{T}i\sigma_2\xi_1\tilde{\Phi}_5
+\mu_5\tilde{\Phi}_6^{T}i\sigma_2\xi_2\tilde{\Phi}_6
\nonumber \\ &&
+\mu_6\tilde{\Phi}_5^{T}i\sigma_2\xi_3\tilde{\Phi}_5
+\mu_7(\tilde{\Phi}_1^{T}i\sigma_2\xi_4
+\omega^2\tilde{\Phi}_2^{T}i\sigma_2\xi_5
+\omega\tilde{\Phi}_3^{T}i\sigma_2\xi_6)\tilde{\Phi}_5
\nonumber \\ &&
+\mu_8(\tilde{\Phi}_1^{T}i\sigma_2\xi_4
+\omega\tilde{\Phi}_2^{T}i\sigma_2\xi_5
+\omega^2\tilde{\Phi}_3^{T}i\sigma_2\xi_6)\tilde{\Phi}_6+h.c..
\end{eqnarray}
In the above two equations, all $\lambda$ parameters are dimensionless,
$\mu$ parameters have mass dimensions and $m_{5,6}^2$ have mass-square
dimensions. We choose $m_{5,6}^2\sim (100~{\rm GeV})^2$ so that the
VEVs of $\Phi_{5,6}$ can be of the order of VEVs of other Higgs
doublets. We suppress the $\mu$
parameters in order to conceive small VEVs for Higgs triplets. Due to
this suppression, one can notice that $\langle V_1^\prime(\Phi,\xi)\rangle$
is very small in comparison to $\langle V_0^\prime(\Phi)\rangle$.

As stated previously, terms in $V_0^\prime(\Phi)$ can affect
the vacuum alignment of Higgs doublets $\Phi_{1,2,3}$. To study these
effects, we minimize $V_0(\Phi)+
V_0^\prime(\Phi)$ with respect to $\phi_1^0,\phi_2^0,\phi_3^0$ and
thereby we get three relations. We solve these relations by demanding
$\langle\phi_i^0\rangle=v$ for $i=1,2,3$. Thereafter we get the following
relations.
\begin{eqnarray}
&&\left[m^2+(3\lambda_1+2\lambda_3+\lambda_4+\lambda_4^*)|v|^2
+(\lambda_1^{(4)}+\lambda_3^{(1)})|v_5|^2
+(\lambda_1^{(5)}+\lambda_3^{(2)})|v_6|^2\right]v
\nonumber \\ &&
+2\lambda_4^{(3)^*}v_6v_5v^*=0,
\nonumber \\ &&
(\lambda_2^{(2)^*}+\lambda_4^{(4)})v_6v_5^*v+\lambda_4^{(1)^*}v^*v_5^2=0,
\quad
(\lambda_2^{(2)}+\lambda_4^{(4)^*})v_6^*v_5v+\lambda_4^{(2)^*}v^*v_6^2=0.
\label{eq:higgsalin}
\end{eqnarray}
Here, $\langle\phi_5^0\rangle=v_5$ and $\langle\phi_6^0\rangle=v_6$. By
solving the unknown parameters in the above three relations, the vacuum
alignment for the Higgs doublets $\Phi_{1,2,3}$ can be achieved. Now, the
VEVs of $\Phi_{5,6}$ should satisfy the following relations.
\begin{eqnarray}
&&\left[m_5^2+3(\lambda_1^{(4)}+\lambda_3^{(1)})|v|^2+\lambda_1^{(1)}|v_5|^2
+(\lambda_1^{(3)}+\lambda_2^{(1)})|v_6|^2\right]v_5
+3\lambda_4^{(3)}v^2v_6^*=0,
\nonumber \\ &&
\left[m_6^2+3(\lambda_1^{(5)}+\lambda_3^{(2)})|v|^2+\lambda_1^{(2)}|v_6|^2
+(\lambda_1^{(3)}+\lambda_2^{(1)})|v_5|^2\right]v_6
+3\lambda_4^{(3)}v^2v_5^*=0.
\label{eq:56vevs}
\end{eqnarray}
As stated before, we take $m^2,m_{5,6}^2\sim(100~{\rm GeV})^2$ so that
the VEVs for Higgs doublets can be chosen to be around 100 GeV. As a result
of this, relations in Eqs. (\ref{eq:higgsalin}) and (\ref{eq:56vevs}) can be
solved for the unknown $\lambda$ parameters, which can be
${\cal O}(1)$.

The VEVs of Higgs triplets can be found after minimizing the potential
$V_1(\Phi,\xi)+V_1^\prime(\Phi,\xi)$. Expressions for these are
given below.
\begin{eqnarray}
&&\left[m_1^2+3\lambda_5^{(1)}|v|^2+\lambda_5^{(11)}|v_5|^2
+\lambda_5^{(15)}|v_6|^2\right]v_1^\prime
+(\lambda_5^{(19)}+\lambda_6^{(18)^*})v_5^*v_6v_3^\prime
\nonumber \\ &&
+(\lambda_5^{(20)^*}+\lambda_6^{(17)})v_5v_6^*v_2^\prime
-3\mu_1^*v^2-\mu_4^*v_6v_5=0,
\label{v1p}
\\ &&
\left[m_2^2+3\lambda_5^{(2)}|v|^2+\lambda_5^{(12)}|v_5|^2
+\lambda_5^{(16)}|v_6|^2\right]v_2^\prime
+(\lambda_5^{(20)}+\lambda_6^{(17)^*})v_5^*v_6v_1^\prime
\nonumber \\ &&
+(\lambda_5^{(21)^*}+\lambda_6^{(19)})v_5v_6^*v_3^\prime
-\mu_5^*v_6^2=0,
\label{eq:v2p} \\ &&
\left[m_3^2+3\lambda_5^{(3)}|v|^2+\lambda_5^{(13)}|v_5|^2
+\lambda_5^{(17)}|v_6|^2\right]v_3^\prime
+(\lambda_5^{(19)^*}+\lambda_6^{(18)})v_5v_6^*v_1^\prime
\nonumber \\ &&
+(\lambda_5^{(21)}+\lambda_6^{(19)^*})v_5^*v_6v_2^\prime
-\mu_6^*v_5^2=0,
\label{eq:v3p} \\ &&
\left[m_0^2+3\lambda_5^{(4)}|v|^2+\lambda_5^{(14)}|v_5|^2
+\lambda_5^{(18)}|v_6|^2+(\lambda_5^{(22)}+\lambda_6^{(20)^*})v_5^*v_6
+(\lambda_5^{(22)^*}+\lambda_6^{(20)})v_5v_6^*\right]v_4^\prime
\nonumber \\ &&
+(\lambda_5^{(9)}+\lambda_5^{(10)^*})|v|^2v_5^\prime
+(\lambda_5^{(9)^*}+\lambda_5^{(10)})|v|^2v_6^\prime
-\mu^*v^2-\mu_7^*vv_5-\mu_8^*vv_6=0,
\\ &&
\left[m_0^2+3\lambda_5^{(4)}|v|^2+\lambda_5^{(14)}|v_5|^2
+\lambda_5^{(18)}|v_6|^2+\omega(\lambda_5^{(22)}+\lambda_6^{(20)^*})v_5^*v_6
+\omega^2(\lambda_5^{(22)^*}+\lambda_6^{(20)})v_5v_6^*\right]v_5^\prime
\nonumber \\ &&
+(\lambda_5^{(9)}+\lambda_5^{(10)^*})|v|^2v_6^\prime
+(\lambda_5^{(9)^*}+\lambda_5^{(10)})|v|^2v_4^\prime
-\mu^*v^2-\omega\mu_7^*vv_5-\omega^2\mu_8^*vv_6=0,
\\ &&
\left[m_0^2+3\lambda_5^{(4)}|v|^2+\lambda_5^{(14)}|v_5|^2
+\lambda_5^{(18)}|v_6|^2+\omega^2(\lambda_5^{(22)}+\lambda_6^{(20)^*})v_5^*v_6
+\omega(\lambda_5^{(22)^*}+\lambda_6^{(20)})v_5v_6^*\right]v_6^\prime
\nonumber \\ &&
+(\lambda_5^{(9)}+\lambda_5^{(10)^*})|v|^2v_4^\prime
+(\lambda_5^{(9)^*}+\lambda_5^{(10)})|v|^2v_5^\prime
-\mu^*v^2-\omega^2\mu_7^*vv_5-\omega\mu_8^*vv_6=0.
\label{v6p}
\end{eqnarray}
From Eqs. (\ref{eq:v2p}) and (\ref{eq:v3p}), we can notice that the
VEVs for $\xi_2$ and $\xi_3$ can be non-zero due to the contribution from
$\mu$ parameters.
In fact, using Eqs. (\ref{v1p}) $-$ (\ref{v6p}), one can infer
that for ${\cal O}(1)$ $\lambda$ parameters,
all the VEVs of Higgs triplets can be chosen to be around 0.1 eV by
suppressing the $\mu$ parameters accordingly.

We have shown that all the Higgs triplets can acquire non-zero VEVs, after
adding two additional Higgs doublets to the MW model. Moreover, we have
demonstrated that vacuum alignment of the Higgs doublets $\Phi_{1,2,3}$
can be achieved in this model. Hence, in a scenario like this,
results described in Sec. 2 are valid. As a result of that, in this model,
the neutrino masses can have either NO or IO, and moreover, this model
is compatible with current neutrino oscillation data.

\section{LFV decays}

In this section, we compute the branching ratios for LFV decays in the
scenario where we extend the MW model with the Higgs doublets $\Phi_{5,6}$.
As described in Sec. 1, LFV decays can be of the following two types:
$\ell\to3\ell^\prime$, $\ell\to\ell^\prime\gamma$. In our scenario,
decays of the form $\ell\to3\ell^\prime$
are driven by doubly charged triplet Higgses. On the other hand, decays
of the form $\ell\to\ell^\prime\gamma$ are driven by both doubly and
singly charged scalars of this model. In order to compute the
branching ratios for these decays, one needs to obtain the mass
eigenstates for doubly and singly charged scalars. Below we present
these mass eigenstates.

It is to be noted that neutral fields from doublet Higgses, other than the
standard model Higgs, can also contribute to the above mentioned LFV decays
\cite{ma-raj}.
Most of these decays are suppressed due to smallness of charged lepton Yukawa
couplings. However, there are some decays, whose amplitudes are proportional
to tau Yukawa coupling, can give appreciable contribution \cite{ma-raj},
provided the masses of the neutral fields are low.
To study the contribution of neutral fields to LFV decays in our scenario,
we have to diagonalize the mixing
masses among the neutral fields of the Higgs doublets. It is to remind here
that since the VEVs of triplet Higgses are very small, the mixing between
neutral fields of doublets and triplets can be neglected.
In this work, we assume that the masses
for the above mentioned
neutral fields are high enough that their contribution to LFV decays
are suppressed. In this regard, let us mention that in Ref. \cite{pz}, LFV
decays driven by neutral scalar fields are studied. The work done in
Ref. \cite{pz} is based on some flavor models \cite{pz2} containing $A_4$
symmetry, where neutral flavon fields induce LFV decays.

\subsection{Mass eigenstates of doubly and singly charged scalars}

The doubly charged scalars belong to the Higgs triplets of our model.
The masses for these fields can be obtained from the scalar potential of
this model, which is given in the previous section. Since six triplet
Higgses exist in the model, one can expect mixing masses among the
doubly charged scalars. These mixing masses are given below.
\begin{eqnarray}
V&\ni& (\psi_1^{++})^\dagger X\psi_1^{++}+(\psi_2^{++})^\dagger Y\psi_2^{++},
\label{eq:++} \\ &&
\psi_1^{++}=(\xi_1^{++},\xi_2^{++},\xi_3^{++})^T,\quad
\psi_2^{++}=(\xi_4^{++},\xi_5^{++},\xi_6^{++})^T,
\nonumber \\ &&
X=\left(\begin{array}{ccc}
x_{11}&x_{12}&x_{13}\\
x_{12}^*&x_{22}&x_{23}\\
x_{13}^*&x_{23}^*&x_{33}
\end{array}\right),\quad
Y=\left(\begin{array}{ccc}
y_{11}&y_{12}&y_{13}\\
y_{12}^*&y_{22}&y_{23}\\
y_{13}^*&y_{23}^*&y_{33}
\end{array}\right),
\nonumber \\
&& x_{11}=m_1^2+3(\lambda_5^{(1)}+\lambda_6^{(1)})|v|^2+\lambda_5^{(11)}|v_5|^2
+\lambda_5^{(15)}|v_6|^2,
\nonumber \\ &&
x_{22}=m_2^2+3(\lambda_5^{(2)}+\lambda_6^{(2)})|v|^2+\lambda_5^{(12)}|v_5|^2
+\lambda_5^{(16)}|v_6|^2,
\nonumber \\ &&
x_{33}=m_3^2+3(\lambda_5^{(3)}+\lambda_6^{(3)})|v|^2+\lambda_5^{(13)}|v_5|^2
+\lambda_5^{(17)}|v_6|^2,\quad
x_{12}=(\lambda_5^{(20)^*}+\lambda_6^{(13)})v_5v_6^*,
\nonumber \\ &&
x_{13}=(\lambda_5^{(19)}+\lambda_6^{(14)^*})v^*_5v_6,\quad
x_{23}=(\lambda_5^{(21)^*}+\lambda_6^{(15)})v_5v_6^*,
\nonumber \\ &&
y_{11}=m_0^2+3(\lambda_5^{(4)}+\lambda_6^{(4)})|v|^2+\lambda_5^{(14)}|v_5|^2
+\lambda_5^{(18)}|v_6|^2
\nonumber \\ &&
+[(\lambda_5^{(22)^*}+\lambda_6^{(16)})v_5v_6^*+h.c.],
\nonumber \\ &&
y_{22}=m_0^2+3(\lambda_5^{(4)}+\lambda_6^{(4)})|v|^2+\lambda_5^{(14)}|v_5|^2
+\lambda_5^{(18)}|v_6|^2
\nonumber \\ &&
+[\omega^2(\lambda_5^{(22)^*}+\lambda_6^{(16)})v_5v_6^*+h.c.],
\nonumber \\ &&
y_{33}=m_0^2+3(\lambda_5^{(4)}+\lambda_6^{(4)})|v|^2+\lambda_5^{(14)}|v_5|^2
+\lambda_5^{(18)}|v_6|^2+
\nonumber \\ &&
+[\omega(\lambda_5^{(22)^*}+\lambda_6^{(16)})v_5v_6^*+h.c.],
\nonumber \\ &&
y_{12}=(\lambda_5^{(9)}+\lambda_5^{(10)^*}
+\lambda_6^{(9)}+\lambda_6^{(10)})|v|^2,\quad
y_{13}=y_{12}^*,\quad y_{23}=y_{12}.
\end{eqnarray}

From Eq. (\ref{eq:++}) we can notice that there is no mixing between
$\xi_{1,2,3}^{++}$ and $\xi_{4,5,6}^{++}$. However, from the quartic
terms of the potential, which are given in Appendix B, there may be
mixing between the above mentioned doubly charged scalars. One can
expect this mixing to be proportional to square of the VEVs of Higgs
triplets, which in our case is very small. Hence, we neglect the above
mentioned mixing. After diagonalizing the matrices $X,Y$ of Eq. (\ref{eq:++}),
we get mass eigenstates for doubly charged scalars, which are defined
below.
\begin{equation}
\xi_{i}^{++}=\sum_{k=1}^3U^{++}_{ik}\xi_{k}^{(m)++},\quad
\xi_{i+3}^{++}=\sum_{k=1}^3V^{++}_{ik}\xi_{k+3}^{(m)++}.
\end{equation}
Here, $i=1,2,3$ and $\xi_{k}^{(m)++}$, where $k=1,\cdots,6$, are the
mass eigenstates of the doubly charged scalars. The unitary matrices
$U^{++},V^{++}$ diagonalize $X,Y$ as
\begin{eqnarray}
&&(U^{++})^\dagger XU^{++}={\rm diag}(M^2_{++(1)},M^2_{++(2)},M^2_{++(3)}),
\\
&&(V^{++})^\dagger YV^{++}={\rm diag}(M^2_{++(4)},M^2_{++(5)},M^2_{++(6)}).
\end{eqnarray}

In analogy to doubly charged scalars, mass eigenstates for singly charged
scalars can be obtained. Singly charged scalars belong to both doublet and
triplet Higgses. In our scenario, due to smallness of VEVs of Higgs triplets,
we can neglect the mixing among singly charged scalars between doublet and
triplet Higgses. Singly charged scalars of doublet Higgses can drive LFV
decays $\ell\to\ell^\prime\gamma$ through charged lepton Yukawa couplings.
One can expect this contribution to be small unless these decays are induced
by tau Yukawa coupling. To simplyfy our analysis we assume
the masses for the singly charged scalars of doublet Higgses
are high enough that their contribution to LFV decays is suppressed.
As a result of this, in this model, the
above mentioned LFV decays are dominantly driven by singly charged scalars
of triplet Higgses. For these reasons, below we present the mass eigenstates
for singly charged scalars from triplet Higgses. These scalars can have
mixing masses, which can be written as
\begin{eqnarray}
V&\ni& (\psi_1^{+})^\dagger X^\prime\psi_1^{+}
+(\psi_2^{+})^\dagger Y^\prime\psi_2^{+},
\label{eq:+} \\ &&
\psi_1^{+}=(\xi_1^{+},\xi_2^{+},\xi_3^{+})^T,\quad
\psi_2^{+}=(\xi_4^{+},\xi_5^{+},\xi_6^{+})^T,
\nonumber \\ &&
X^\prime=\left(\begin{array}{ccc}
x^\prime_{11}&x^\prime_{12}&x^\prime_{13}\\
x_{12}^{\prime^*}&x^\prime_{22}&x^\prime_{23}\\
x_{13}^{\prime^*}&x_{23}^{\prime^*}&x^\prime_{33}
\end{array}\right),\quad
Y^\prime=\left(\begin{array}{ccc}
y_{11}^\prime&y_{12}^\prime&y_{13}^\prime\\
y_{12}^{\prime^*}&y_{22}^\prime&y_{23}^\prime\\
y_{13}^{\prime^*}&y_{23}^{\prime^*}&y_{33}^\prime
\end{array}\right),
\nonumber \\ &&
x_{11}^\prime=x_{11}-\frac{3}{2}\lambda_6^{(1)}|v|^2,\quad
x_{22}^\prime=x_{22}-\frac{3}{2}\lambda_6^{(2)}|v|^2,\quad
x_{33}^\prime=x_{33}-\frac{3}{2}\lambda_6^{(3)}|v|^2,
\nonumber \\ &&
x_{12}^\prime=x_{12}-\frac{1}{2}(\lambda_6^{(13)}-\lambda_6^{(17)})v_5v_6^*,
\quad
x_{13}^\prime=x_{13}-\frac{1}{2}(\lambda_6^{(14)^*}-\lambda_6^{(18)^*})
v_5^*v_6,
\nonumber \\ &&
x_{23}^\prime=x_{23}-\frac{1}{2}(\lambda_6^{(15)}-\lambda_6^{(19)})v_5v_6^*,
\nonumber \\ &&
y_{11}^\prime=y_{11}-\frac{3}{2}\lambda_6^{(4)}|v|^2
-\frac{1}{2}[(\lambda_6^{(16)^*}-\lambda_6^{(20)^*})v_5^*v_6+h.c.],
\nonumber \\ &&
y_{22}^\prime=y_{22}-\frac{3}{2}\lambda_6^{(4)}|v|^2
-\frac{1}{2}[\omega(\lambda_6^{(16)^*}-\lambda_6^{(20)^*})v_5^*v_6+h.c.],
\nonumber \\ &&
y_{33}^\prime=y_{33}-\frac{3}{2}\lambda_6^{(4)}|v|^2
-\frac{1}{2}[\omega^2(\lambda_6^{(16)^*}-\lambda_6^{(20)^*})v_5^*v_6+h.c.],
\nonumber \\ &&
y_{12}^\prime=y_{12}-\frac{1}{2}(\lambda_6^{(9)}+\lambda_6^{(10)}
-\lambda_6^{(11)^*}-\lambda_6^{(12)^*})|v|^2,\quad
y_{13}^\prime=(y_{12}^\prime)^*,\quad y_{23}^\prime=y_{12}^\prime.
\end{eqnarray}

From Eq. (\ref{eq:+}), in analogy to doubly charged scalars, we can
notice that there is no mixing between $\xi_{1,2,3}^+$ and $\xi_{4,5,6}^+$
at leading order. Now, we can define the mass eigenstates for singly
charged scalars as
\begin{equation}
\xi_{i}^{+}=\sum_{k=1}^3U^{+}_{ik}\xi_{k}^{(m)+},\quad
\xi_{i+3}^{+}=\sum_{k=1}^3V^{+}_{ik}\xi_{k+3}^{(m)+}.
\label{eq:maeig+}
\end{equation}
Here, $i=1,2,3$ and $\xi_{k}^{(m)+}$, where $k=1,\cdots,6$, are the
mass eigenstates of singly charged scalars. The unitary matrices
$U^{+},V^{+}$ diagonalize $X^\prime,Y^\prime$ as
\begin{eqnarray}
&&(U^{+})^\dagger X^\prime U^{+}={\rm diag}(M^2_{+(1)},M^2_{+(2)},M^2_{+(3)}),
\\
&&(V^{+})^\dagger Y^\prime V^{+}={\rm diag}(M^2_{+(4)},M^2_{+(5)},M^2_{+(6)}).
\end{eqnarray}

\subsection{Branching ratios of $\ell\to3\ell^\prime$}

In this subsection, we compute the branching ratios for decays of the
form $\ell\to3\ell^\prime$. Since these decays are driven by doubly
charged scalars at tree level, we need to obtain couplings between
doubly charged scalars and charged leptons. These couplings are determined
by the Lagrangian of Eq. (\ref{eq:lag}), where all the scalars and fermions
of this Lagrangian are in flavor states. For charged leptons, by applying
the transformations in Eq. (\ref{eq:translep}), we get the corresponding
mass eigenstates. For doubly charged scalars, the mass eigenstates have
been described in the previous subsection. After using the above mentioned
mass eigenstates in Eq. (\ref{eq:lag}), we get the desired couplings needed
for the decays $\ell\to3\ell^\prime$. These are given below.
\begin{eqnarray}
{\cal L}&\ni&-\sum_{k,l=1,k\leq l}^3\ell_k^{(m)^T}C\frac{1-\gamma_5}{2}
\left[\sum_{j=1}^3
f_{1j}^{k,l}\xi_j^{(m)++}+f_{2j}^{k,l}\xi_{j+3}^{(m)++}\right]\ell_l^{(m)},
\nonumber \\ &&
f_{1j}^{1,1}=y_1U_{1j}^{++},\quad
f_{2j}^{1,1}=\frac{y}{3}(V_{1j}^{++}+V_{2j}^{++}+V_{3j}^{++}),\quad
f_{1j}^{2,3}=2f_{1j}^{1,1},\quad f_{2j}^{2,3}=-f_{2j}^{1,1},
\nonumber \\ &&
f_{1j}^{2,2}=y_2U_{2j}^{++},\quad
f_{2j}^{2,2}=\frac{y}{3}(V_{1j}^{++}+\omega^2V_{2j}^{++}+\omega V_{3j}^{++}),
\quad
f_{1j}^{1,3}=2f_{1j}^{2,2},\quad f_{2j}^{1,3}=-f_{2j}^{2,2},
\nonumber \\ &&
f_{1j}^{3,3}=y_3U_{3j}^{++},\quad
f_{2j}^{3,3}=\frac{y}{3}(V_{1j}^{++}+\omega V_{2j}^{++}+\omega^2V_{3j}^{++}),
\quad
f_{1j}^{1,2}=2f_{1j}^{3,3},\quad f_{2j}^{1,2}=-f_{2j}^{3,3}.
\nonumber \\
\label{eq:l3l}
\end{eqnarray}
Here, $C$ is the charge conjugation matrix and $\ell^{(m)}_j$
is a mass eigenstate of charged lepton. From the above equation, we
can notice that some
of the couplings between doubly charged scalars and charged leptons are
related to one another. This is a result due to $A_4$ symmetry of the model.
This result has implications
on the branching ratios of the decays of the form $\ell\to3\ell^\prime$.
We will explain these implications shortly later.

Using the couplings in Eq. (\ref{eq:l3l}), we compute the branching
ratios for $\ell\to3\ell^\prime$, after neglecting the masses of final
state charged leptons. Branching ratios for $\tau$ decays are
found to be
\begin{equation}
{\rm Br}(\tau\to\bar{\ell}_i\ell_j\ell_k)=\frac{S}{32G_F^2}
\left|\sum_{n=1}^3\frac{(f_{1n}^{j,k})^*f_{1n}^{i,3}}{M^2_{++(n)}}
+\frac{(f_{2n}^{j,k})^*f_{2n}^{i,3}}{M^2_{++(n+3)}}\right|^2
{\rm Br}(\tau\to\mu\bar{\nu}\nu).
\label{eq:taubr}
\end{equation}
Here, $G_F$ is the Fermi constant and
${\rm Br}(\tau\to\mu\bar{\nu}\nu)=0.1739$ \cite{pdg}. Moreover, the indices
$i,j,k=1,2$ are for electron and muon fields. $S=1(2)$ if
$\ell_j\neq\ell_k(\ell_j=\ell_k)$. In the above equation, one should
use $f_{1n}^{j,k}=f_{1n}^{k,j}$ and $f_{2n}^{j,k}=f_{2n}^{k,j}$.
These relations follow from the Lagrangian of Eq. (\ref{eq:l3l}).
The branching ratio for $\mu\to3e$ is
\begin{equation}
{\rm Br}(\mu\to\bar{e}ee)=\frac{1}{16G_F^2}
\left|\sum_{n=1}^3\frac{(f_{1n}^{1,1})^*f_{1n}^{1,2}}{M^2_{++(n)}}
+\frac{(f_{2n}^{1,1})^*f_{2n}^{1,2}}{M^2_{++(n+3)}}\right|^2.
\label{eq:mubr}
\end{equation}
In this work, we have assumed ${\rm Br}(\mu\to e\bar{\nu}\nu)=100\%$.

As stated before, some relations exist among the couplings in the
Lagrangian of Eq. (\ref{eq:l3l}). An implication of this is there can
exist relations among branching ratios for some decays. From
Eq. (\ref{eq:taubr}), we can see that
\begin{equation}
{\rm Br}(\tau\to\bar{e}ee)={\rm Br}(\tau\to\bar{\mu}\mu\mu).
\label{eq:br1}
\end{equation}
From Eqs. (\ref{eq:taubr}) and (\ref{eq:mubr}), after assuming degenerate
values for $M^2_{++(4)},M^2_{++(5)},M^2_{++(6)}$, we get
\begin{equation}
{\rm Br}(\tau\to\bar{\mu}e\mu)=2{\rm Br}(\mu\to\bar{e}ee)
{\rm Br}(\tau\to \mu\bar{\nu}\nu).
\label{eq:br2}
\end{equation}
On the other hand, in the limit where the masses for all doubly charged
scalars are degenerate, Eqs. (\ref{eq:taubr}) and (\ref{eq:mubr}) imply that the
branching ratios for following decays go to zero: $\tau\to\bar{e}ee$,
$\tau\to\bar{e}e\mu$, $\tau\to\bar{\mu}\mu\mu$, $\tau\to\bar{\mu}e\mu$,
$\mu\to\bar{e}ee$. Relations among the branching ratios described in
Eqs. (\ref{eq:br1}) and (\ref{eq:br2}) are due to the $A_4$ symmetry
of our model. In the work of Ref. \cite{pz}, which is based on $A_4$ symmetry,
a similar kind of relations among
various branching ratios for $\ell\to3\ell^\prime$ have been derived. Since
the flavor models considered in Ref. \cite{pz} are different from our model,
the relations for branching ratios given in Ref. \cite{pz} are different
from Eqs. (\ref{eq:br1}) and (\ref{eq:br2}). We can notice here that
searching for LFV decays in experiments can distinguish various flavor models.
Moreover, these searches can give some hints about $A_4$ symmetry.

Among various decays of the type $\ell\to3\ell^\prime$, branching
ratio for $\mu\to\bar{e}ee$ is severely constrained. From experiments,
we have ${\rm Br}(\mu\to\bar{e}ee)<1.0\times 10^{-12}$ \cite{sindrum}.
In order to satisfy this constraint in our work, we study the branching
ratio of $\mu\to\bar{e}ee$. From Eq. (\ref{eq:mubr}), we can see that
${\rm Br}(\mu\to\bar{e}ee)$ depends on masses of doubly charged scalars
and on couplings between doubly charged scalars and charged leptons.
These couplings, which can be seen from Eq. (\ref{eq:l3l}), depend on
neutrino Yukawa couplings and the unitary matrices
which diagonalize the mixing masses for doubly charged scalars. Hence, the
masses for doubly charged scalars and the above mentioned unitary matrices
are determined from the parameters of the scalar potential. On the other hand,
the neutrino Yukawa couplings are determined from the VEVs of Higgs triplets
and neutrino oscillation observables. This fact can be seen from Eqs.
(\ref{eq:vev}), (\ref{yuexp}) and (\ref{eq:eps}). From these equations,
one can notice that the neutrino Yukawa couplings depend on $\theta_{13}$
and $\theta_{23}$, but not on $\theta_{12}$.

As described above, ${\rm Br}(\mu\to\bar{e}ee)$, in our work, depend on
neutrino oscillation observables, VEVs of Higgs triplets and parameters of
scalar potential. It is interesting to see the variation of
${\rm Br}(\mu\to\bar{e}ee)$ in terms of neutrino oscillation observables.
Hence, we have fixed VEVs of Higgs triplets and parameters of scalar
potential to some specific values in our analysis. The details of our
analysis have been described below.

To simplify our numerical analysis, we take all the
independent Higgs triplet VEVs to be same as $v_T$. It is to remind here
that the VEVs $v_4^\prime,v_5^\prime$ are not independent. It is
discussed in Sec. 2 that the Yukawa coupling $y$ can be determined in terms
of $v_4^\prime$ or $v_5^\prime$. As a result of this, from Eq.
(\ref{eq:case}), we can see that in case I(II)
$v_4^\prime(v_5^\prime)$ is independent parameter. As for the masses of
doubly charged scalars, they
are determined after diagonalizing the $X,Y$ matrices of Eq. (\ref{eq:++}).
Since there are several $\lambda$ parameters exist in $X,Y$, for the
sake of illustration, we choose all these $\lambda$ parameters to be 0.1.
We take the mass-square parameters of $X,Y$ as $m_1^2=m_2^2=m_3^2=m_0^2
=(850~{\rm GeV})^2$. We have taken the VEVs for doublet Higgses as
$v=v_5=v_6=174/\sqrt{5}$ GeV. With the above mentioned parameters, we
have found the masses for all doubly charged scalars to be slightly
above 850 GeV. These mass values for doubly charged scalars satisfy
the lower bound on them, which is obtained by the LHC experiment \cite{ll}.

In Figs. 1 and 2, we have given the plots for branching ratios of
$\mu\to\bar{e}ee$, in the cases of NO and IO respectively.
\begin{figure}[!h]
\begin{center}

\includegraphics[width=3.0in]{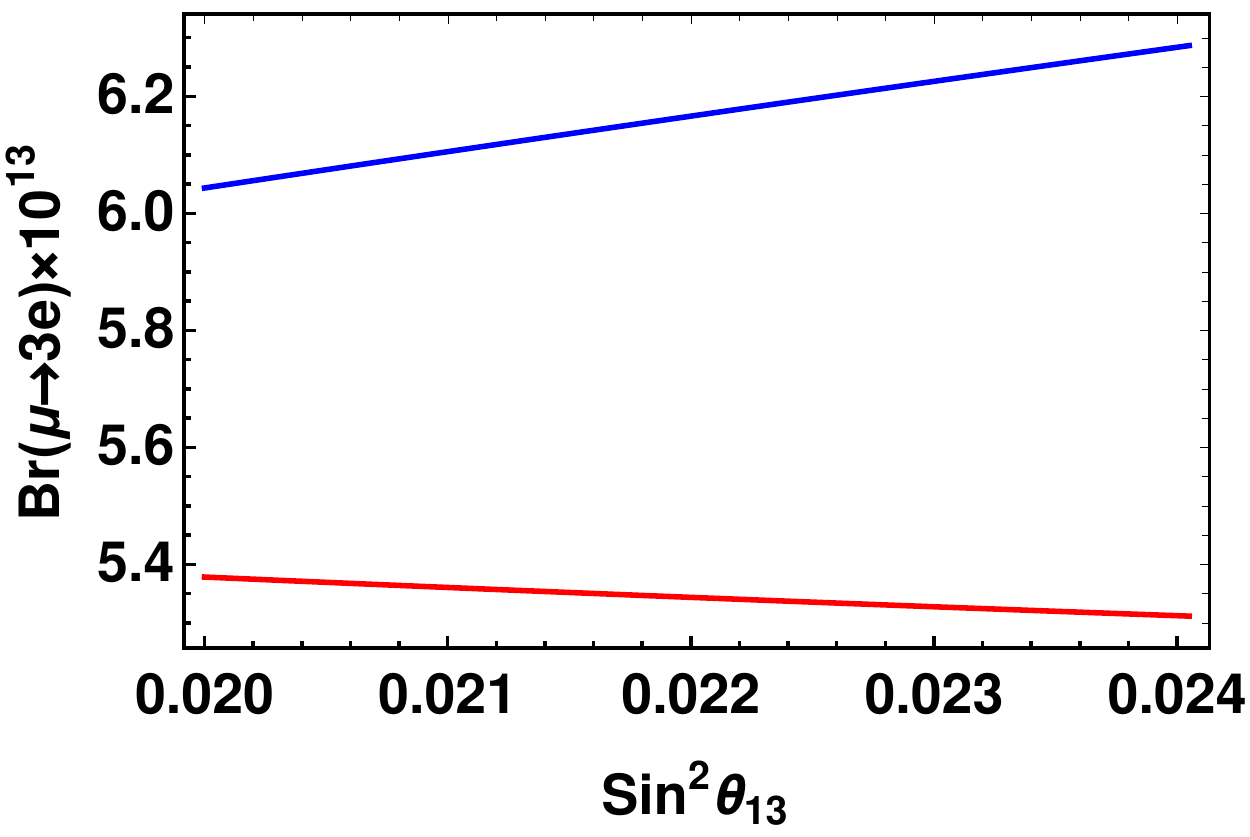}
\includegraphics[width=3.0in]{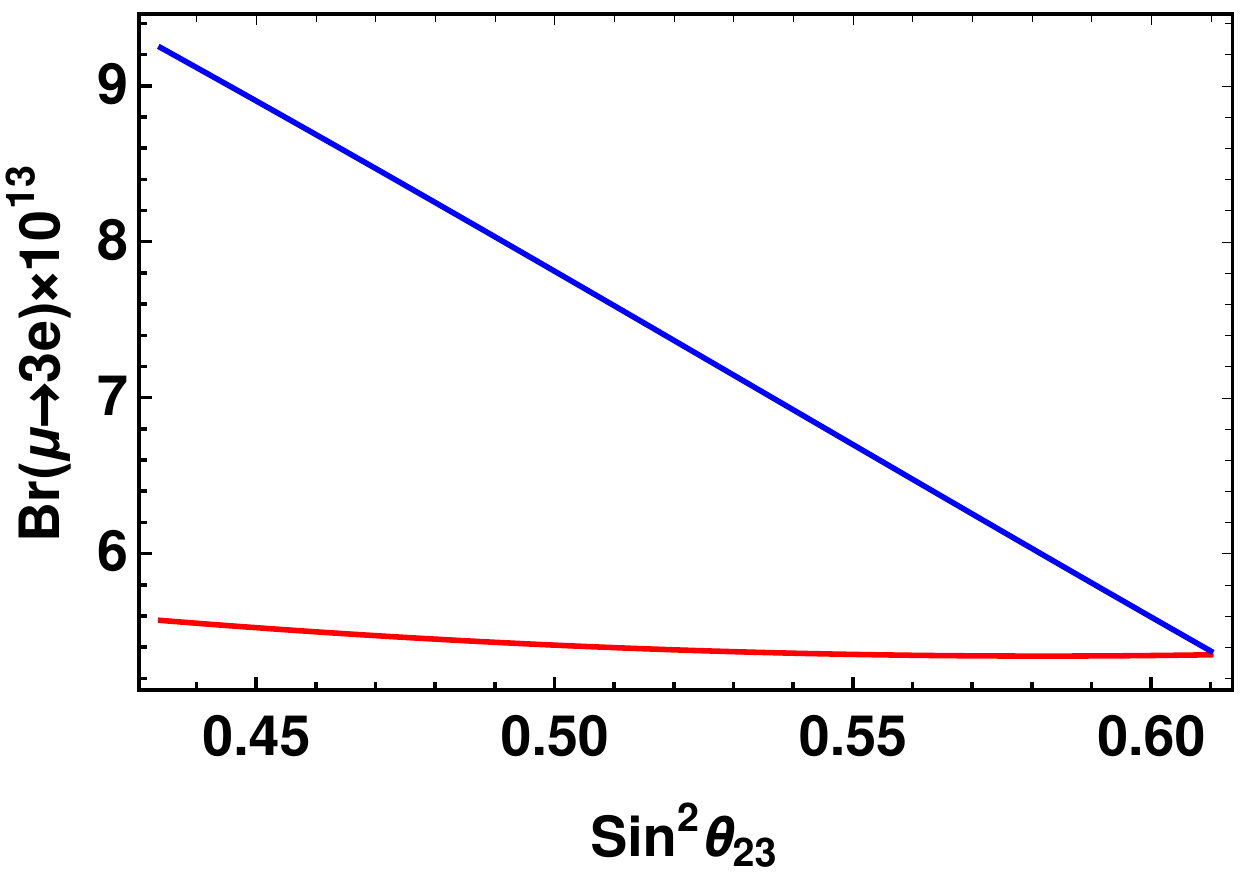}
\includegraphics[width=3.0in]{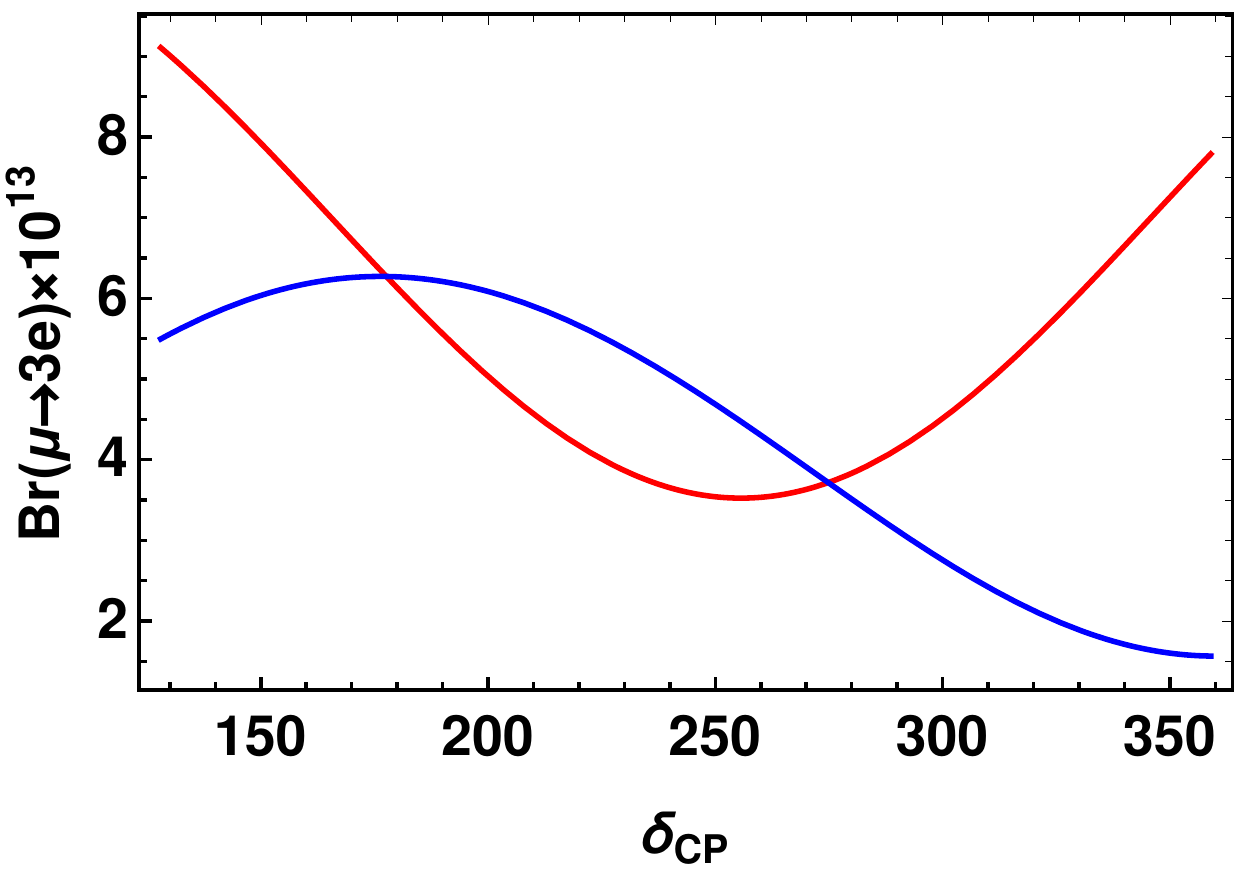}

\end{center}
\caption{Branching ratios for $\mu\to\bar{e}ee$ in the case of NO. Here,
red and blue lines are for the cases I and II respectively. $\delta_{CP}$
is expressed in degrees. In these plots, $v_T=0.08$ eV. In the top-left
plot, $\sin^2\theta_{23}$ and $\delta_{CP}$ are fixed to the best fit values
of Table 2. In the top-right plot, $\sin^2\theta_{13}$ and $\delta_{CP}$
are fixed to the best fit values of Table 2. In the bottom plot,
$\sin^2\theta_{13}$ and $\sin^2\theta_{23}$ are fixed to the best fit
values of Table 2. In all these plots, lightest neutrino mass is taken to
be zero and the other neutrino masses are computed from Eqs.
(\ref{eq:msquare}) and
(\ref{eq:numval}). For details related to masses of doubly charged scalars,
see the text.}
\end{figure}
\begin{figure}[!h]
\begin{center}

\includegraphics[width=3.0in]{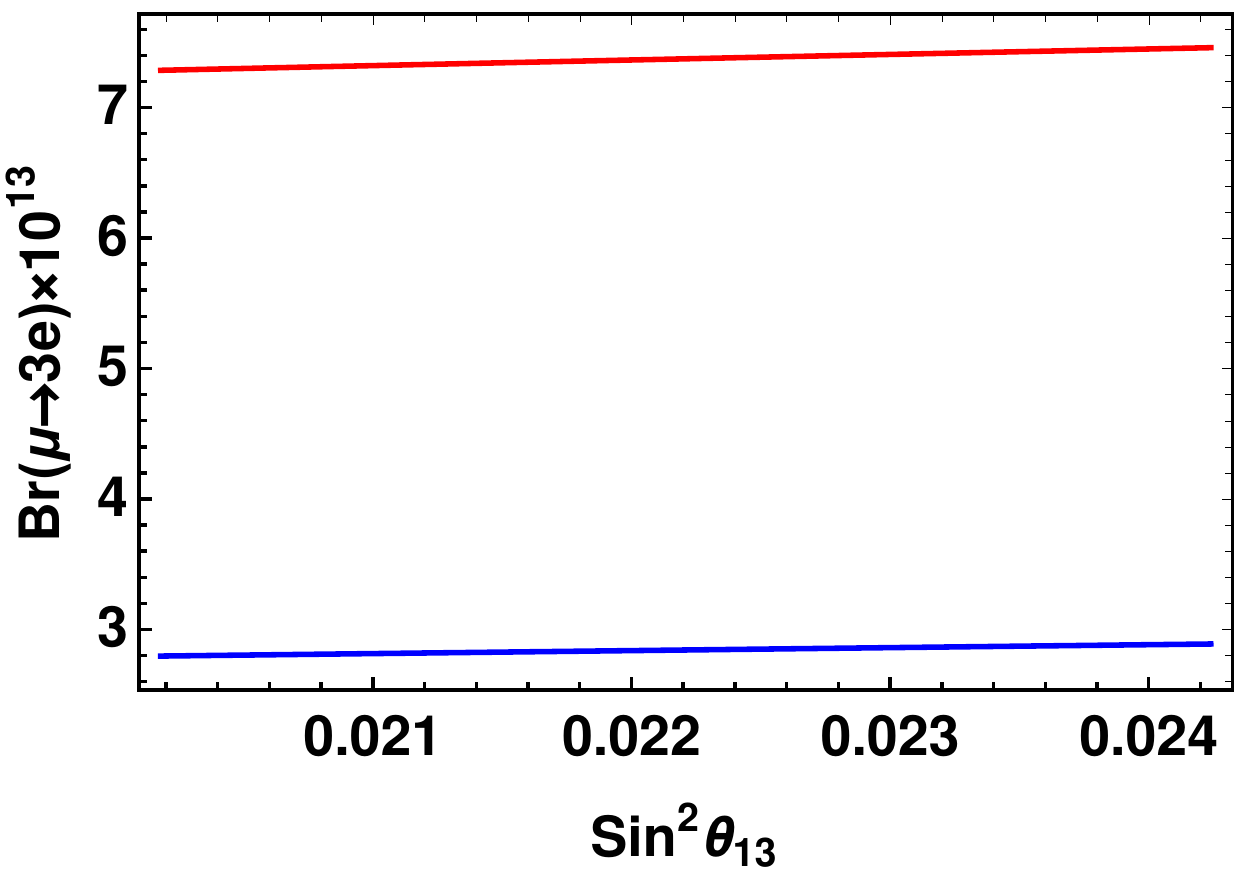}
\includegraphics[width=3.0in]{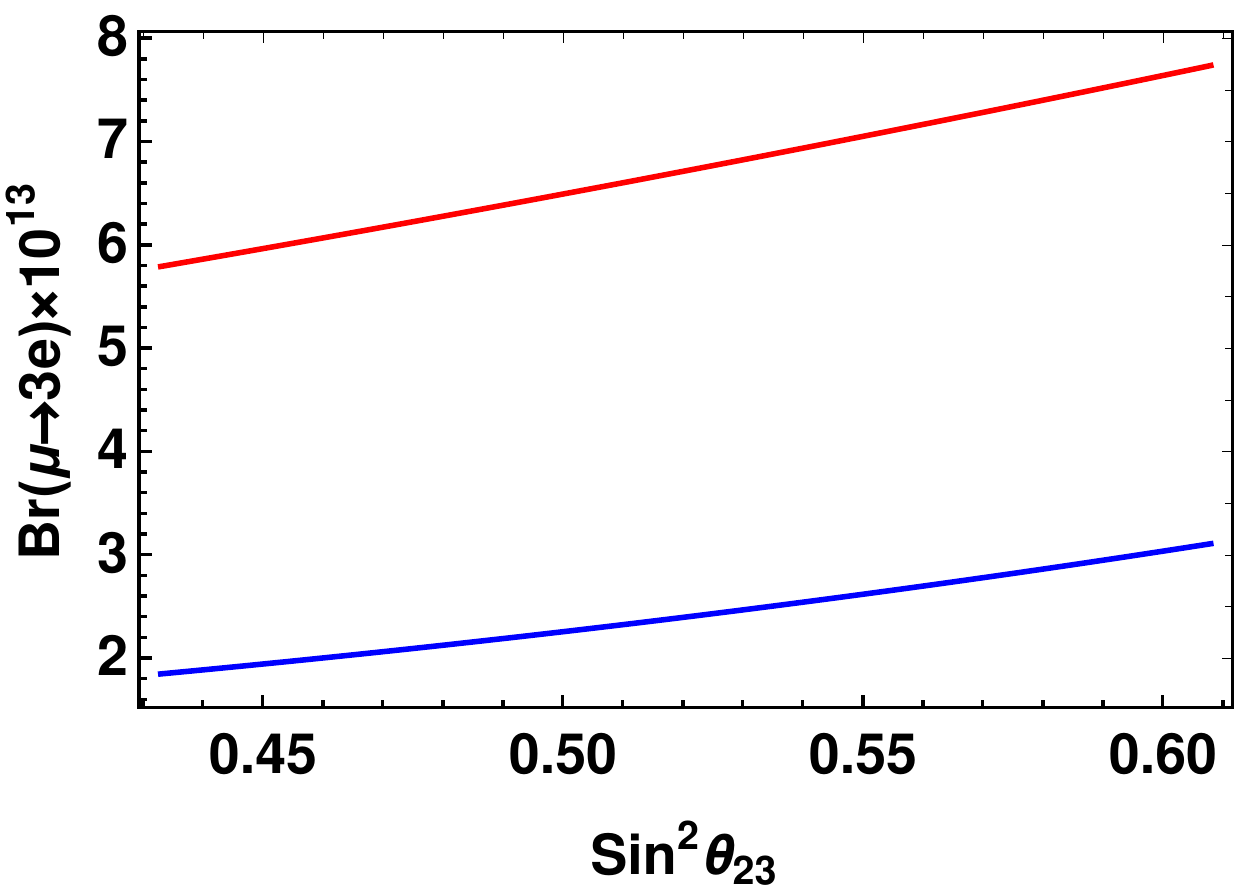}
\includegraphics[width=3.0in]{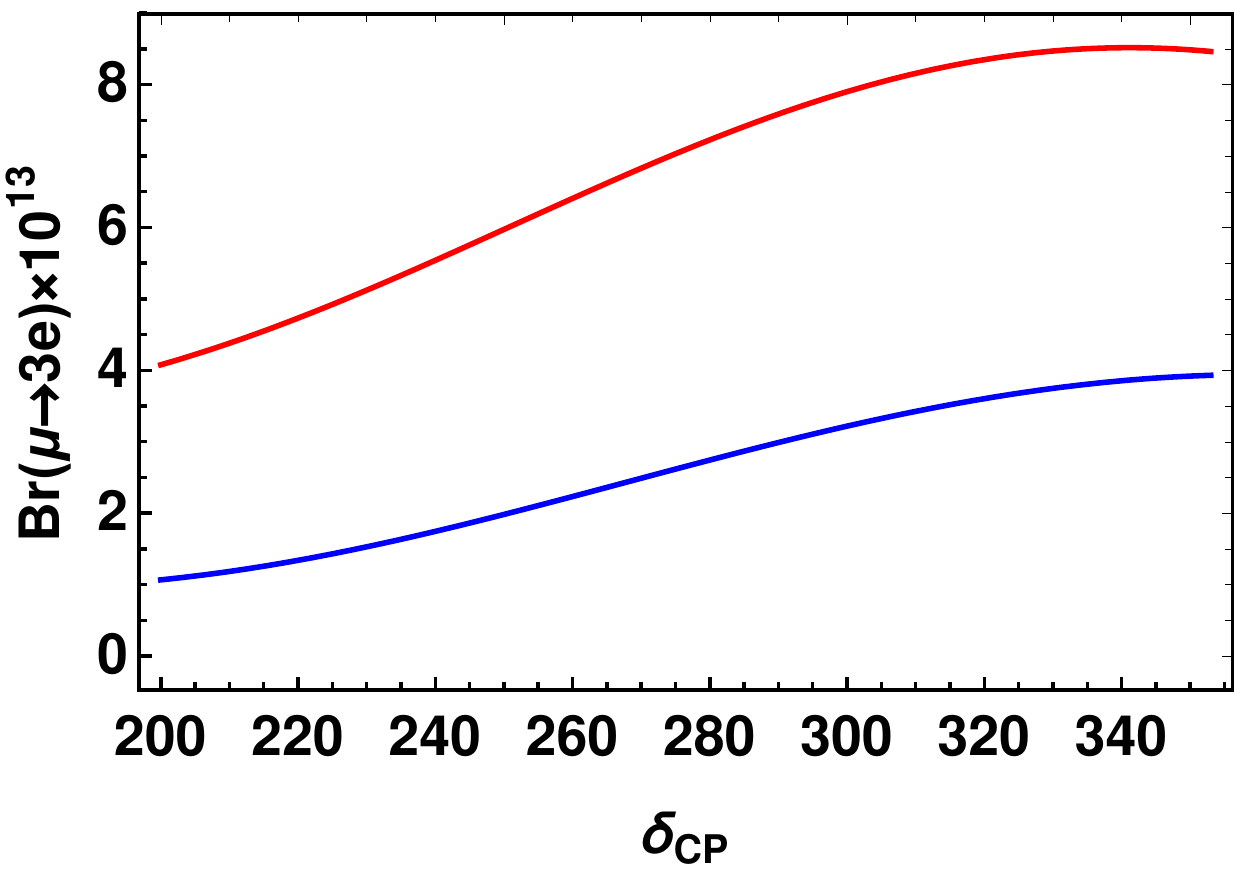}

\end{center}
\caption{Branching ratios for $\mu\to\bar{e}ee$ in the case of IO. Here,
red and blue lines are for the cases I and II respectively. $\delta_{CP}$
is expressed in degrees. In these plots, $v_T=0.14$ eV. In the top-left
plot, $\sin^2\theta_{23}$ and $\delta_{CP}$ are fixed to the best fit values
of Table 2. In the top-right plot, $\sin^2\theta_{13}$ and $\delta_{CP}$
are fixed to the best fit values of Table 2. In the bottom plot,
$\sin^2\theta_{13}$ and $\sin^2\theta_{23}$ are fixed to the best fit
values of Table 2. In all these plots, lightest neutrino mass is taken to
be zero and the other neutrino masses are computed from Eqs.
(\ref{eq:msquare}) and
(\ref{eq:numval}). For details related to masses of doubly charged
scalars, see the text.}
\end{figure}
As already described above, the neutrino Yukawa couplings of our model,
up to the leading order, do not depend on the mixing angle
$\theta_{12}$. Hence, in Figs. 1 and 2, ${\rm Br}(\mu\to\bar{e}ee)$ is
plotted against $\sin^2\theta_{13}$, $\sin^2\theta_{23}$ and $\delta_{CP}$.
The allowed ranges and best fit values for the neutrino mixing angles
and $\delta_{CP}$, which are used in this work, are tabulated in Table 2.
\begin{table}[!h]
\centering
\begin{tabular}{|c|c c|} \hline
parameter & best fit & 3$\sigma$ range \\\hline
$\sin^2\theta_{13}/10^{-2}$ (NO) & 2.200 & 2.000 - 2.405 \\
$\sin^2\theta_{13}/10^{-2}$ (IO) & 2.225 & 2.018 - 2.424 \\
$\sin^2\theta_{23}/10^{-1}$ (NO) & 5.74 & 4.34 - 6.10 \\
$\sin^2\theta_{23}/10^{-1}$ (IO) & 5.78 & 4.33 - 6.08 \\
$\delta_{CP}/{\rm o}$ (NO) & 194 & 128 - 359 \\
$\delta_{CP}/{\rm o}$ (IO) & 284 & 200 - 353 \\\hline
\end{tabular}
\caption{Values of the neutrino oscillation parameters \cite{glo-fit},
which are used in this work.}
\end{table}
In Figs. 1 and 2, in the plot between ${\rm Br}(\mu\to\bar{e}ee)$ and
$\sin^2\theta_{13}$, we have fixed the best fit values for
$\sin^2\theta_{23}$ and $\delta_{CP}$, which are given in Table 2.
Similar kind of things have been done in other plots of Figs. 1 and 2.
In the plots of both these figures, we have taken the lightest neutrino
mass to be zero and the other neutrino masses are computed from Eqs.
(\ref{eq:msquare}) and (\ref{eq:numval}). In Figs. 1 and 2,
we have taken $v_T$ to be 0.08 eV and 0.14 eV respectively.
If we decrease $v_T$ below than the above mentioned values,
the value for ${\rm Br}(\mu\to\bar{e}ee)$ may exceed the experimental
limit on this in the plots of Figs. 1 and 2. One can notice, in each plot
of these figures we get two lines, which is due to the fact that the
Yukawa coupling $y$ can be determined either in terms of $v_4^\prime$
or $v_5^\prime$. Depending on our choice of free parameter between
$v_4^\prime$ and $v_5^\prime$, the branching
ratio for $\mu\to\bar{e}ee$ can be different in this model, which is
evident from Figs. 1 and 2. Which of these two choices is true is
something we may tell after measuring the branching ratio for
this decay in experiments.

\subsection{Branching ratios of $\ell\to\ell^\prime\gamma$}

As stated before, decays of the form $\ell\to\ell^\prime\gamma$ are
driven by both doubly and singly charged triplet scalars. Interaction
terms between doubly charged scalars and charged leptons, which are given in
Eq. (\ref{eq:l3l}), drive $\ell\to\ell^\prime\gamma$ at 1-loop level.
In addition to this contribution, singly charged triplet scalars interacting
with charged leptons and neutrinos also contribute to
$\ell\to\ell^\prime\gamma$ at 1-loop level. To obtain these interaction
terms, which involve singly charged scalars, we use the transformations for
left-handed charged leptons and neutrinos of
Eqs. (\ref{eq:translep}) and (\ref{eq:trans}) in Eq. (\ref{eq:lag}),
apart from using Eq. (\ref{eq:maeig+}).
As a result of this, we get the following interaction terms for singly
charged triplet scalars with charged leptons and neutrinos.
\begin{eqnarray}
{\cal L}&\ni&-\sum_{j,k=1}^3\nu_k^{(m)^T}C\frac{1-\gamma_5}{2}\left[
g_{jk}^{1,1}\xi_j^{(m)+}+g_{jk}^{1,2}\xi_{j+3}^{(m)+}\right]\ell_1^{(m)}
\nonumber \\ &&
-\sum_{j,k=1}^3\nu_k^{(m)^T}C\frac{1-\gamma_5}{2}\left[
g_{jk}^{2,1}\xi_j^{(m)+}+g_{jk}^{2,2}\xi_{j+3}^{(m)+}\right]\ell_2^{(m)}
\nonumber \\ &&
-\sum_{j,k=1}^3\nu_k^{(m)^T}C\frac{1-\gamma_5}{2}\left[
g_{jk}^{3,1}\xi_j^{(m)+}+g_{jk}^{3,2}\xi_{j+3}^{(m)+}\right]\ell_3^{(m)}+h.c.,
\nonumber \\ &&
g_{jk}^{1,1}=\sqrt{2}[y_1U^+_{1j}(U_{PMNS})_{1k}
+y_2U^+_{2j}(U_{PMNS})_{3k}+y_3U^+_{3j}(U_{PMNS})_{2k}],
\nonumber \\ &&
g_{jk}^{1,2}=\frac{y}{3\sqrt{2}}[2(V^+_{1j}+V^+_{2j}+V^+_{3j})(U_{PMNS})_{1k}
-(V^+_{1j}+\omega V^+_{2j}+\omega^2V^+_{3j})(U_{PMNS})_{2k}
\nonumber \\ &&
-(V^+_{1j}+\omega^2V^+_{2j}+\omega V^+_{3j})(U_{PMNS})_{3k}],
\nonumber \\ &&
g_{jk}^{2,1}=\sqrt{2}[y_1U^+_{1j}(U_{PMNS})_{3k}
+y_2U^+_{2j}(U_{PMNS})_{2k}+y_3U^+_{3j}(U_{PMNS})_{1k}],
\nonumber \\ &&
g_{jk}^{2,2}=\frac{y}{3\sqrt{2}}[
-(V^+_{1j}+\omega V^+_{2j}+\omega^2V^+_{3j})(U_{PMNS})_{1k}
+2(V^+_{1j}+\omega^2 V^+_{2j}+\omega V^+_{3j})(U_{PMNS})_{2k}
\nonumber \\ &&
-(V^+_{1j}+V^+_{2j}+V^+_{3j})(U_{PMNS})_{3k}],
\nonumber \\ &&
g_{jk}^{3,1}=\sqrt{2}[y_1U^+_{1j}(U_{PMNS})_{2k}
+y_2U^+_{2j}(U_{PMNS})_{1k}+y_3U^+_{3j}(U_{PMNS})_{3k}],
\nonumber \\ &&
g_{jk}^{3,2}=\frac{y}{3\sqrt{2}}[
-(V^+_{1j}+\omega^2 V^+_{2j}+\omega V^+_{3j})(U_{PMNS})_{1k}
-(V^+_{1j}+V^+_{2j}+V^+_{3j})(U_{PMNS})_{2k}
\nonumber \\ &&
+2(V^+_{1j}+\omega V^+_{2j}+\omega^2V^+_{3j})(U_{PMNS})_{3k}].
\label{eq:lnuxi}
\end{eqnarray}
In the above equation, $\nu_k^{(m)}$, where $k=1,2,3,$ are mass eigenstates for
neutrinos.

Using the interaction terms of Eqs. (\ref{eq:l3l})
and (\ref{eq:lnuxi}), the total amplitude for the decay
$\ell\to\ell^\prime\gamma$ can be written as
\begin{equation}
{\cal M}=-\frac{Q_e^2}{24\pi^2}(a_{++}^{\ell,\ell^\prime}
+\frac{1}{8}a_+^{\ell,\ell^\prime})\epsilon_\mu^*(q)
\bar{u}_{\ell^\prime}(p-q)\left[m_{\ell^\prime}\frac{1-\gamma_5}{2}
+m_\ell\frac{1+\gamma_5}{2}\right]i\sigma^{\mu\nu}q_\nu u_\ell(p).
\label{eq:amp}
\end{equation}
Here, $m_{\ell^\prime}$ and $m_\ell$ are masses for the charged leptons
$\ell^\prime$ and $\ell$ respectively. $Q_e$ is the magnitude of charge of
electron. The quantities
$a_{++}^{\ell,\ell^\prime}$ and $a_{+}^{\ell,\ell^\prime}$ depend on
masses of triplet charged scalars and their couplings with leptons. Their forms
are given below.
\begin{eqnarray}
&& a_{++}^{\ell,\ell^\prime}=\sum_{j=1}^3\frac{a_{1j}^{++(\ell,\ell^\prime)}}
{M_{++(j)}^2}+\frac{a_{2j}^{++(\ell,\ell^\prime)}}{M_{++(j+3)}^2},\quad
a_{+}^{\ell,\ell^\prime}=\sum_{j=1}^3\frac{a_{1j}^{+(\ell,\ell^\prime)}}
{M_{+(j)}^2}+\frac{a_{2j}^{+(\ell,\ell^\prime)}}{M_{+(j+3)}^2},
\nonumber \\ &&
a_{nj}^{++(\mu,e)}=(f_{nj}^{1,1})^*f_{nj}^{1,2}
+\frac{1}{2}(f_{nj}^{1,3})^*f_{nj}^{2,3}+(f_{nj}^{1,2})^*f_{nj}^{2,2},\quad
n=1,2,
\nonumber \\ &&
a_{1j}^{+(\mu,e)}=\sum_{k=1}^3(g_{jk}^{1,1})^*g_{jk}^{2,1},\quad
a_{2j}^{+(\mu,e)}=\sum_{k=1}^3(g_{jk}^{1,2})^*g_{jk}^{2,2},
\nonumber \\ &&
a_{nj}^{++(\tau,\mu)}=\frac{1}{2}(f_{nj}^{1,2})^*f_{nj}^{1,3}
+(f_{nj}^{2,2})^*f_{nj}^{2,3}+(f_{nj}^{2,3})^*f_{nj}^{3,3},\quad n=1,2,
\nonumber \\ &&
a_{1j}^{+(\tau,\mu)}=\sum_{k=1}^3(g_{jk}^{2,1})^*g_{jk}^{3,1},\quad
a_{2j}^{+(\tau,\mu)}=\sum_{k=1}^3(g_{jk}^{2,2})^*g_{jk}^{3,2},
\nonumber \\ &&
a_{nj}^{++(\tau,e)}=(f_{nj}^{1,1})^*f_{nj}^{1,3}
+\frac{1}{2}(f_{nj}^{1,2})^*f_{nj}^{2,3}+(f_{nj}^{1,3})^*f_{nj}^{3,3},\quad
n=1,2,
\nonumber \\ &&
a_{1j}^{+(\tau,e)}=\sum_{k=1}^3(g_{jk}^{1,1})^*g_{jk}^{3,1},\quad
a_{2j}^{+(\tau,e)}=\sum_{k=1}^3(g_{jk}^{1,2})^*g_{jk}^{3,2}.
\end{eqnarray}
Using the amplitude in Eq. (\ref{eq:amp}), we find the branching ratios
for the decays of the form $\ell\to\ell^\prime\gamma$, where we have neglected
the mass of $\ell^\prime$. Expressions for these are given below.
\begin{eqnarray}
&&{\rm Br}(\tau\to\ell^\prime\gamma)=\frac{\alpha}{12\pi G_F^2}
\left|a_{++}^{\tau,\ell^\prime}+\frac{1}{8}a_+^{\tau,\ell^\prime}\right|^2
{\rm Br}(\tau\to\mu\bar{\nu}\nu),
\nonumber \\
&&{\rm Br}(\mu\to e\gamma)=\frac{\alpha}{12\pi G_F^2}
\left|a_{++}^{\mu,e}+\frac{1}{8}a_+^{\mu,e}\right|^2.
\label{eq:loopd}
\end{eqnarray}
Here, $\alpha=\frac{Q_e^2}{4\pi}$ and $\ell^\prime=e,\mu$.

In the previous
subsection, we have shown in Eqs. (\ref{eq:br1}) and (\ref{eq:br2})
that branching ratios for different decays of the form $\ell\to3\ell^\prime$
can relate to each other. We have explained that this is due to an
implication of $A_4$ symmetry, under which the couplings of doubly
charged scalars can relate to one another. We have found that even for
the decays of the form $\ell\to\ell^\prime\gamma$, there can exist relations
among branching ratios of different decays, under some particular
conditions. If $M_{++(j)}^2,M_{+(j)}^2$ are degenerate for $j=4,5,6$,
from Eq. (\ref{eq:loopd}) we get
\begin{equation}
{\rm Br}(\tau\to\mu\gamma)={\rm Br}(\mu\to e\gamma)
{\rm Br}(\tau\to\mu\bar{\nu}\nu)
\label{eq:br3}
\end{equation}
On the other hand, if $M_{++(j-3)}^2,M_{++(j)}^2,M_{+(j)}^2$ are
degenerate for $j=4,5,6$, we get
\begin{equation}
{\rm Br}(\tau\to\mu\gamma)={\rm Br}(\tau\to e\gamma)={\rm Br}(\mu\to e\gamma)
{\rm Br}(\tau\to\mu\bar{\nu}\nu)
\label{eq:br4}
\end{equation}
We can also notice that in the limit where all the masses of doubly and
singly charged
scalar triplets are degenerate, the branching ratios in Eq. (\ref{eq:loopd})
go to zero. Verifying the relations of Eqs. (\ref{eq:br3}) and
(\ref{eq:br4}) in experiments can give some hints about $A_4$ symmetry
of this model. Notice here that, in a related work of Ref. \cite{pz},
similar kind of relations among the branching ratios for the decays
$\ell\to\ell^\prime\gamma$ have been given.

Among the various decays of the form $\ell\to\ell^\prime\gamma$,
branching ratio for $\mu\to e\gamma$ is severely constrained and
we have ${\rm Br}(\mu\to e\gamma)<4.2\times10^{-13}$ \cite{meg}. From the
expression given for ${\rm Br}(\mu\to e\gamma)$ in Eq. (\ref{eq:loopd}),
one can see that this depends on the masses and couplings of both doubly
and singly charged triplet Higgses. The couplings of doubly and singly
charged triplets are given in Eqs. (\ref{eq:l3l}) and (\ref{eq:lnuxi}).
These couplings depend on neutrino Yukawa couplings and also on
parameters of scalar potential. Now,
from the discussion given for the case of ${\rm Br}(\mu\to\bar{e}ee)$, one
can realize that ${\rm Br}(\mu\to e\gamma)$ in our work is determined
by neutrino oscillation observables, VEVs of Higgs triplets and parameters
of the scalar potential. From the same discussion, one can also realize
that ${\rm Br}(\mu\to e\gamma)$ in our work do not depend on the mixing
angle $\theta_{12}$, at the leading order. Since it is interesting to
study variation of ${\rm Br}(\mu\to e\gamma)$ with respect to neutrino
oscillation observables, we have fixed VEVs of Higgs triplets and
parameters of the scalar potential to some specific values, which will
be described below. It should be noticed that both ${\rm Br}(\mu\to e\gamma)$
and ${\rm Br}(\mu\to\bar{e}ee)$ are determined by a common set of
parameters, since doubly charged triplet Higgses contribute to both of
the above observables. In addition to this common set of observables,
${\rm Br}(\mu\to e\gamma)$ is determined by parameters related to singly
charged triplet Higgses.

We have
computed ${\rm Br}(\mu\to e\gamma)$ in our model for the cases of NO and IO,
which are presented in Figs. 3 and 4 respectively.
\begin{figure}[!h]
\begin{center}

\includegraphics[width=3.0in]{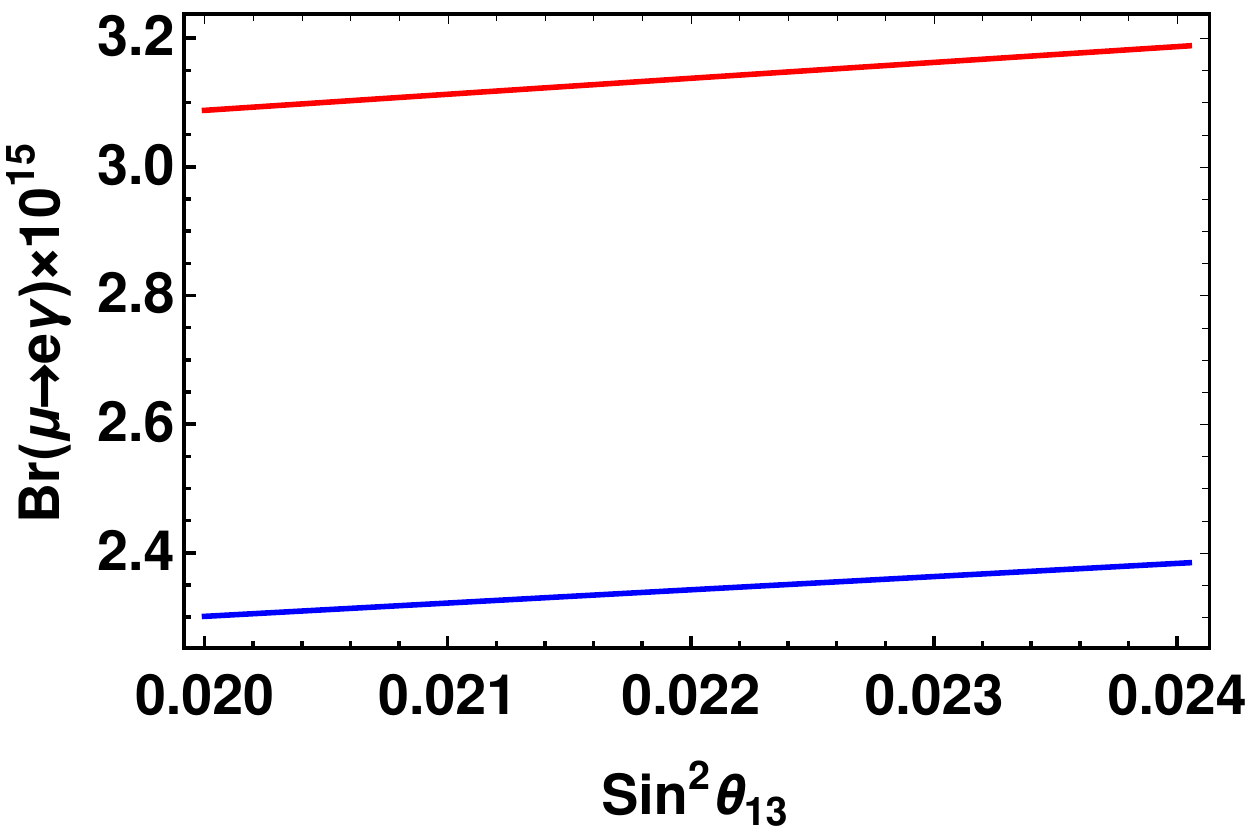}
\includegraphics[width=3.0in]{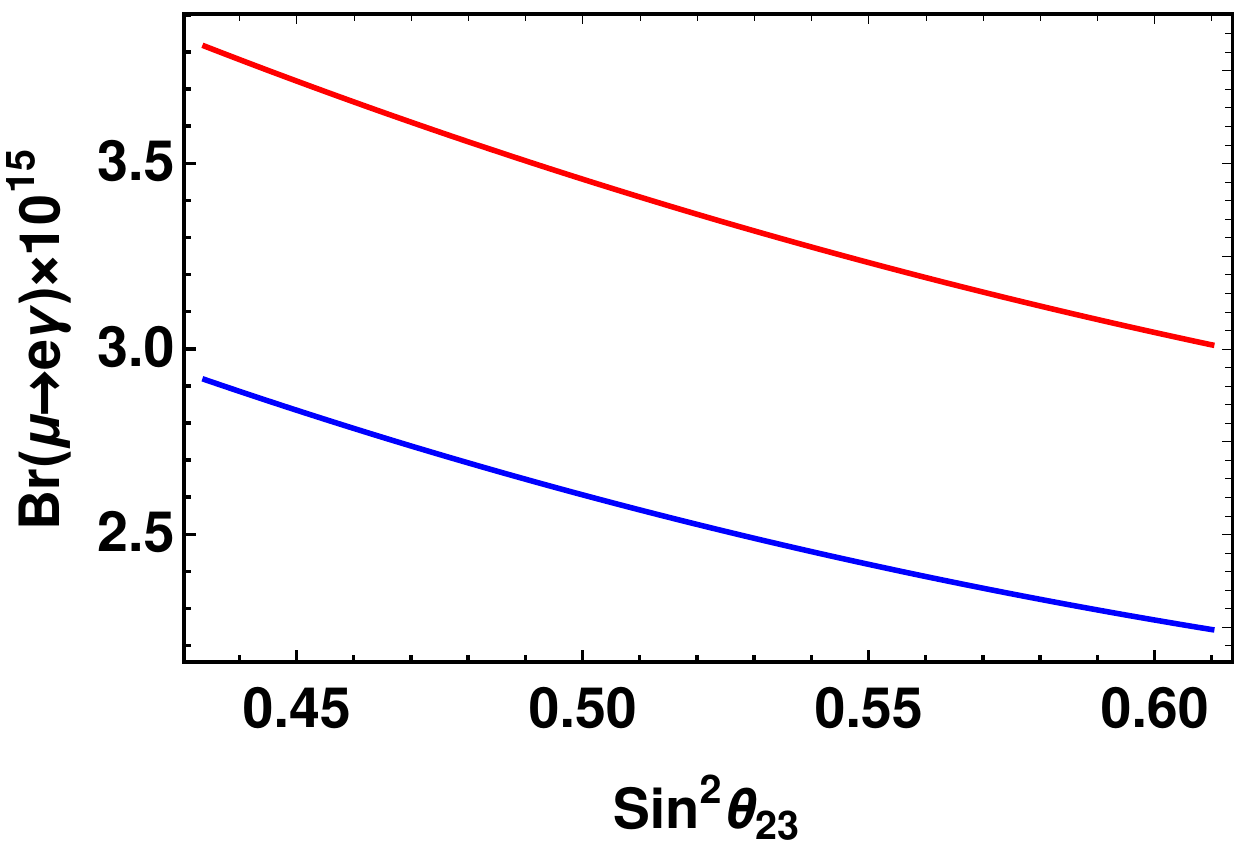}
\includegraphics[width=3.0in]{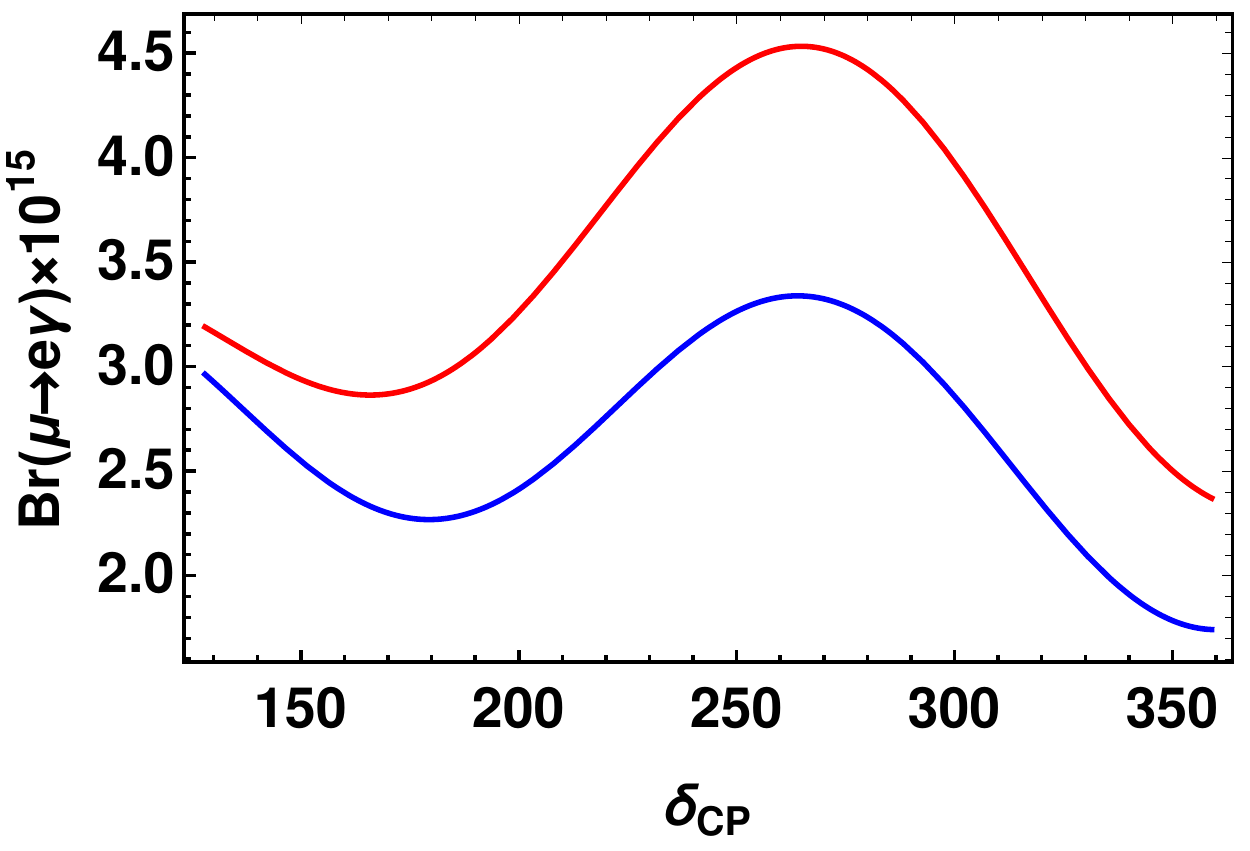}

\end{center}
\caption{Branching ratios for $\mu\to e\gamma$ in the case of NO. Here,
red and blue lines are for the cases I and II respectively. $\delta_{CP}$
is expressed in degrees. In these plots, $v_T=0.08$ eV. The neutrino
oscillation parameters in these plots are taken to be same as for Fig. 1.
For details related to charged triplet scalar masses, see the text.}
\end{figure}
\begin{figure}[!h]
\begin{center}

\includegraphics[width=3.0in]{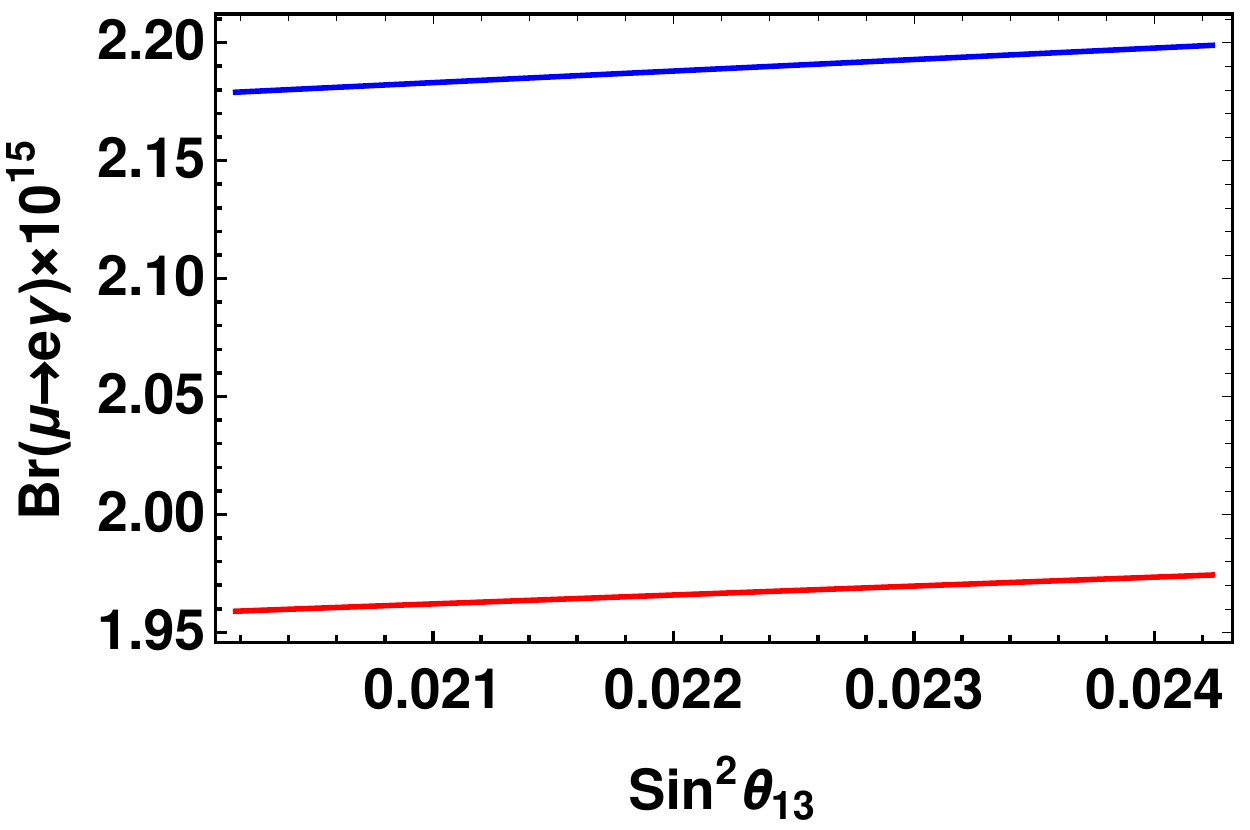}
\includegraphics[width=3.0in]{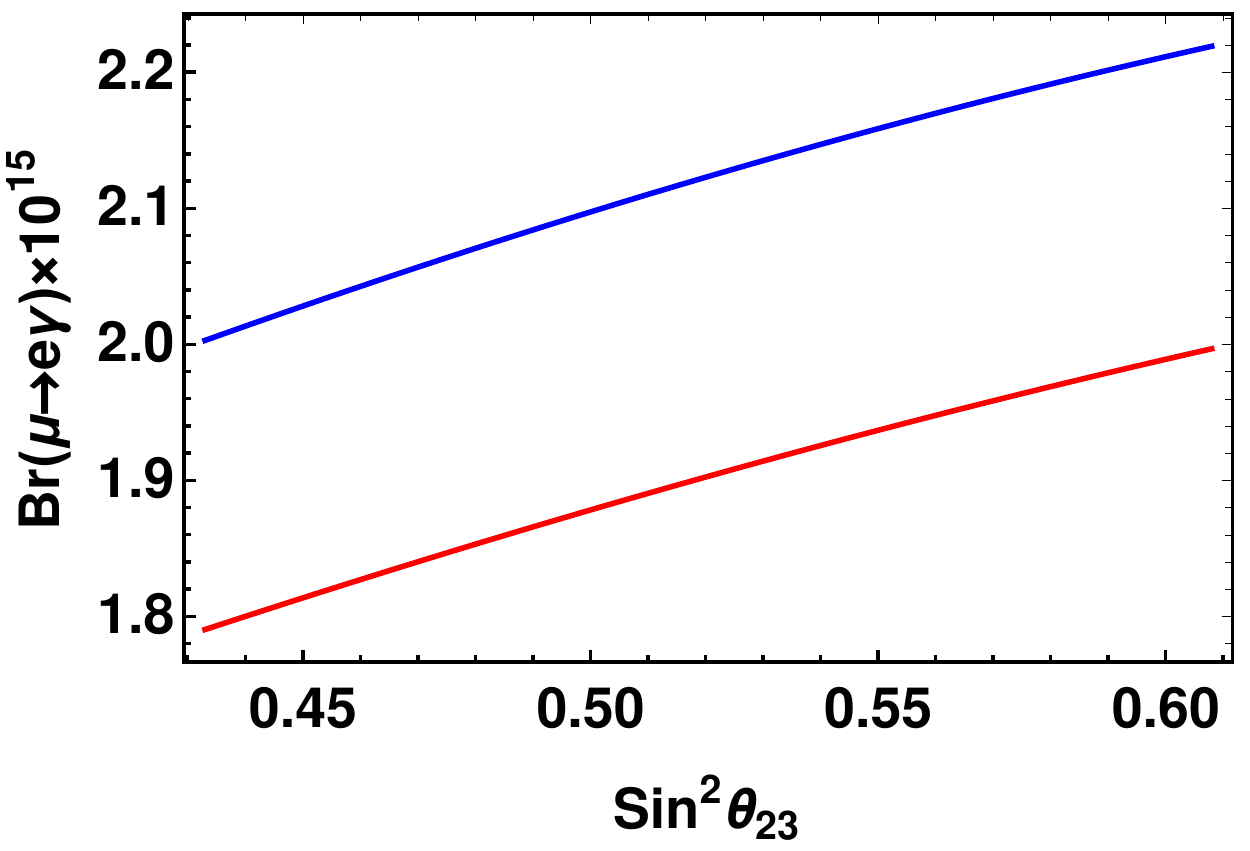}
\includegraphics[width=3.0in]{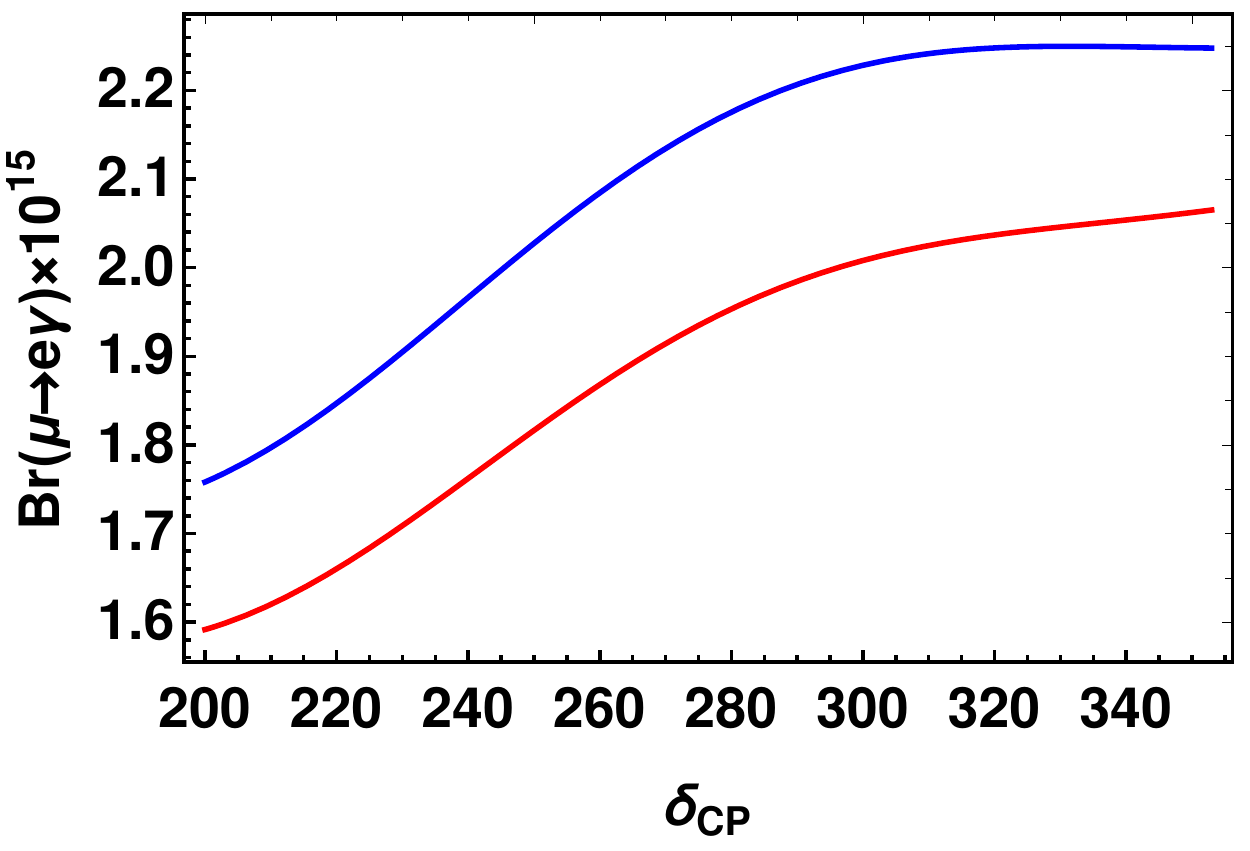}

\end{center}
\caption{Branching ratios for $\mu\to e\gamma$ in the case of IO. Here,
red and blue lines are for the cases I and II respectively. $\delta_{CP}$
is expressed in degrees. In these plots, $v_T=0.14$ eV. The neutrino
oscillation parameters in these plots are taken to be same as for Fig. 2.
For details related to charged triplet scalar masses, see the text.}
\end{figure}
While computing the ${\rm Br}(\mu\to e\gamma)$, we have used the same set
of parameters which are described for the computation of
${\rm Br}(\mu\to\bar{e}ee)$. Now
the additional parameters which govern the decay $\mu\to e\gamma$ are
due to the singly charged triplet scalar fields. The masses and couplings
of these singly charged scalars are determined after diagonalizing the
mass matrices for these, which are given in Eq. (\ref{eq:+}). There is a
common set of parameters in the mass matrices for singly and
doubly charged scalar fields. This common set of parameters is same as what we
have used for the computation of ${\rm Br}(\mu\to\bar{e}ee)$. The additional
$\lambda$ parameters in the
mass matrices of singly charged triplet scalars are taken to be 0.1
in this analysis. As a result of this, the masses for both doubly and
singly charged triplets are slightly above 850 GeV.
After using the above mentioned parameters for the computation of
${\rm Br}(\mu\to e\gamma)$,
from Figs. 3 and 4, we can see that the branching ratio for this decay
is around $10^{-15}$. This value of branching ratio is two orders lower
than that for $\mu\to\bar{e}ee$, whose results can be seen from Figs. 1 and 2.
The reason for this suppression in the
branching ratio is due to the fact that the decays $\mu\to e\gamma$ and
$\mu\to\bar{e}ee$ take place at 1-loop and tree level respectively. As
a result of this, a loop suppression factor of $\alpha\sim10^{-2}$
exist in the ${\rm Br}(\mu\to e\gamma)$, which gives the above mentioned
suppression.

In the upcoming MEG II experiment,
the sensitivity to probe ${\rm Br}(\mu\to e\gamma)$ is around $10^{-14}$
\cite{meg2}. Hence, the parameter region of Figs. 3 and 4 may not be
reachable in the upcoming MEG II experiment. We can get
${\rm Br}(\mu\to e\gamma)\sim 10^{-14}$ in this analysis, by decreasing
the values of either $v_T$ or the masses for charged triplet fields. However,
in such cases the branching ratio for $\mu\to\bar{e}ee$ may exceed the
experimental limit on this decay. Moreover, it is to be noted that we
have chosen the parametric values of $m_1^2=m_2^2=m_3^2=m_0^2
=(850~{\rm GeV})^2$ in such a way that the doubly charged scalar fields
have masses above 850 GeV. The current stringent lower bound on
the doubly charged scalar mass is around 850 GeV \cite{ll}.
By decreasing the values for above mentioned mass-square
parameters, one needs to ensure that the lower bound on the doubly
charged scalar masses are satisfied. One can do a detailed study on the
above mentioned topic, nevertheless, we can notice that probing LFV
decays in experiments can reveal something about our model, which is based
on the MW model. Finally, in each plot of the Figs. 3 and 4, the two lines
correspond to the choice of the free parameter between $v_4^\prime$ and
$v_5^\prime$, which is described around Eq. (\ref{eq:case}).
Depending on this choice of parameter, the branching
ratio for $\mu\to e\gamma$ can be different. After this decay is observed
in experiments, by matching the theoretical formula for
${\rm Br}(\mu\to e\gamma)$ with the observed data, we may tell
which of the above mentioned parameters can be chosen free.

It is mentioned previously that contribution from the neutral scalar fields
to the LFV decays is neglected in this work. Even after including
this contribution, it is still an interest to know the results about LFV
decays, in the limit
where the masses of these fields are heavy enough that the contribution
can be neglected. In this work, we have analyzed the above mentioned case.
On the other hand, depending on the masses and coupling strengths of these
neutral scalar fields, the results mentioned in this work
can be altered. It is worth to study this contribution, however,
it is stated that only the neutral scalars which interact with tau lepton
may give appreciable contribution. Before studying this contribution,
one has to diagonalize the mixing masses among the neutral scalar fields,
which is an involved work and we postpone it to future.

\section{Future directions and phenomenology of our model}

The model presented in this work contains additional scalar fields which are
five Higgs doublets and six Higgs triplets. After the electroweak symmetry
breaking, the following fields remain in the theory: six doubly charged
scalars, ten singly charged scalars, twenty one neutral scalars. One of
these neutral scalars can be identified as the Higgs boson, which is discovered
in the LHC. All the above mentioned scalars have gauge interactions. Hence,
it is possible to produce them at the LHC, and after production, they can
subsequently decay into standard model fields via their Yukawa or gauge
interactions. So the model presented in this work can be tested at the LHC.
We have shown that this model can make certain predictions in LFV decays,
which are given in Eqs. (\ref{eq:br1}), (\ref{eq:br2}), (\ref{eq:br3})
and (\ref{eq:br4}). Among these, testing the LFV relation in Eq.
(\ref{eq:br1}) is the best way to check this model in experiments, since
this relation is independent on the assumptions made on the masses of charged
scalars.

From the context of LFV decays, the model presented in this work can be
distinguished from the original MW model. Our model is an extension
of MW model with additional Higgs doublets $\Phi_{5,6}$. Hence, by putting
$\langle\Phi_{5,6}\rangle=0$ in our results of
LFV, one can get corresponding results in the MW model. After using
$\langle\Phi_{5,6}\rangle=0$ in the mixing mass matrices of doubly and
singly charged triplets, which are given in Sec. 4.1, one can notice that
doubly and singly charged scalars of $\xi_{1,2,3}$ are already in
mass eigenstates. On the other hand, doubly and singly charged scalars of
$\xi_{4,5,6}$ can mix non-trivially. As a result of this, in the MW model,
LFV decays are driven by only the doubly and singly charged scalars of
$\xi_{4,5,6}$, in contrast to the fact that these decays are driven by all
charged triplet Higgses in our model. Hence, the rate of LFV decays in the
MW model can be different from that in our model. This can be one source
to distinguish our model from the MW model in experiments.
Another source to distinguish our model from other $A_4$ symmetry models
is the study of collider implications in the scalar sector.

From the plots of Figs. 1 to 4, we can see that the LFV decays in our
work depend on neutrino oscillation observables. However, due to large
number of parameters in our model, we have simplified the
numerical analysis by choosing some specific values for the parameters
in the scalar potential. Hence, the plots in Figs. 1 to 4 are for
some specific benchmark points of our model, where we have taken all
$\lambda$ parameter to 0.1. An extensive numerical analysis
on LFV decays in our model is still possible.
Since in our model,
neutrinos are Majorana particles, the neutrino oscillation observables
can get additional constraints due to neutrino-less double beta decay.
From the non-observation of this decay, upper bounds have been set on the
effective Majorana mass $m_{ee}$ \cite{pdg}, which is expressed in terms
of neutrino masses and elements of the first row of $U_{PMNS}$. The most
stringent upper bound on $m_{ee}$ is 61 $-$ 165 meV \cite{kzen}. Using this
bound on $m_{ee}$, allowed regions for LFV decays in our work can be
studied. Apart from the above mentioned bounds, precision electroweak
observables \cite{pdg} can also give additional constraints on the model.

The singly and doubly charged scalars of our model can drive $H\to\gamma
\gamma$ at 1-loop level. Here, $H$ is a neutral scalar of our model, which
represents Higgs boson of standard model. The decay rate for $H\to\gamma
\gamma$ in our model depends on the tri-linear couplings of $H$ with
singly and doubly charged scalars. These couplings are determined by the
parameters of the scalar potential of our model. Since the signal strength
for $H\to\gamma\gamma$ at the LHC \cite{pdg} agrees with the standard model
prediction for Higgs boson, there can be additional constraints on the
above mentioned tri-linear couplings in our model.

In Sec. 3.2, we have given the minimization conditions for the doublet
and triplet Higgses of our model. These conditions can represent a possible
minimum for the scalar potential of our model. This minimum may or may not
be a global minimum of our scalar potential. We may expect additional
conditions to be imposed on the parameters of the scalar potential in order
to make this minimum to be global. For related studies in this direction,
see Refs. \cite{xu}.

In this work, we have studied mixing pattern in lepton sector by introducing
additional Higgs doublets and triplets. It is interesting to know about
masses and mixing pattern of quarks in our framework with $A_4$ symmetry.
In this direction, in Refs. \cite{qmix}, breaking of $A_4$ symmetry is
suggested for obtaining realistic mixing pattern in quark and lepton
sectors. Following these ideas, one can study quark masses and mixing
pattern in our model.

\section{Conclusions}

In this work, we have considered the MW model \cite{ma-weg}, where the
mixing pattern in neutrino sector is explained with three Higgs doublets,
six Higgs triplets and with the additional symmetry $A_4$. The
VEVs of Higgs triplets play a part in explaining the neutrino mixing pattern,
apart from the fact that the
VEVs of Higgs doublets should be same in order to diagonalize the charged
lepton mass matrix. To study the pattern of VEVs of scalar fields of the
MW model, in this work,
we have constructed the invariant scalar potential of this model. After
minimizing this scalar potential, we have found that among the six Higgs
triplets two of them acquire zero VEVs. As a result of this, after using
the results from the diagonalization procedure of our previous work
\cite{prwo}, we have found that the neutrino mixing angles cannot be
consistently explained. In order to see if we can get a consistent picture
with the diagonalization procedure of our previous work \cite{prwo},
we have added two additional Higgs doublets to the
MW model. Thereafter, we have shown that all the Higgs triplets acquire
non-zero VEVs and the current neutrino oscillation data can be explained
in this model. After adding extra Higgs doublets to the model,
we have demonstrated
that enough parameter space exist, where the above mentioned vacuum
alignment of Higgs doublets can be achieved.

To study some phenomenological consequences of the model under
consideration, we have computed branching ratios for the
LFV decays of the form $\ell\to3\ell^\prime$ and $\ell\to\ell^\prime\gamma$.
We have found that $A_4$ symmetry of this model can bring some relations
among the
couplings between charged triplet scalars and lepton fields. As a result of
this, relations can exist among branching ratios for different decays.
Relation shown in Eq. (\ref{eq:br1}) is independent of any assumption
on the masses of charged triplet scalars. However, relations in Eqs.
(\ref{eq:br2}), (\ref{eq:br3}) and (\ref{eq:br4}) are valid under
some assumptions made on the masses of charged triplet scalars.
Apart from this, branching ratios for the LFV decays in our
work depend on the neutrino mixing angles
$\theta_{13}$ and $\theta_{23}$ and also on the $CP$ violating Dirac
phase $\delta_{CP}$. We have plotted branching ratios for these decays
in both the cases of NO and IO. From these plots, we have found that the
choice of free parameters among the VEVs of
Higgs triplets can have implications on the branching ratios for the
LFV decays of this model.

\section*{Appendix A: Product rules of $A_4$ symmetry}

The discrete symmetry $A_4$ has 12 elements which constitute the following
4 irreducible representations: $\underline{1}$, $\underline{1}^\prime$,
$\underline{1}^{\prime\prime}$, $\underline{3}$. Product rules for these
irreducible representations are
\begin{eqnarray}
&&\underline{1}^\prime\times\underline{1}^\prime=
\underline{1}^{\prime\prime},\quad
\underline{1}^{\prime\prime}\times\underline{1}^{\prime\prime}=
\underline{1}^{\prime},\quad
\underline{1}^{\prime}\times\underline{1}^{\prime\prime}=
\underline{1},
\nonumber \\ &&
\underline{1}^\prime\times\underline{3}=\underline{3},\quad
\underline{1}^{\prime\prime}\times\underline{3}=\underline{3},\quad
\underline{3}\times\underline{3}=\underline{1}+\underline{1}^\prime+
\underline{1}^{\prime\prime}+\underline{3}_1+\underline{3}_2.
\end{eqnarray}
Let $(x_1,x_2,x_3)$ and $(y_1,y_2,y_3)$ be two triplets under $A_4$.
Then we have \cite{a4rules}
\begin{eqnarray}
&&\underline{1}=x_1y_1+x_2y_2+x_3y_3,\quad
\underline{1}^\prime=x_1y_1+\omega^2x_2y_2+\omega x_3y_3,\quad
\underline{1}^{\prime\prime}=x_1y_1+\omega x_2y_2+\omega^2x_3y_3,
\nonumber \\ &&
\underline{3}_1=(x_2y_3,x_3y_1,x_1y_2),\quad
\underline{3}_2=(x_3y_2,x_1y_3,x_2y_1).
\end{eqnarray}
Let $u\sim\underline{1}^\prime$ and $v\sim\underline{1}^{\prime\prime}$.
Then we have \cite{a4rules}
\begin{equation}
\underline{1}^\prime\times\underline{3}=u(x_1,\omega x_2,\omega^2x_3),\quad
\underline{1}^{\prime\prime}\times\underline{3}=v(x_1,\omega^2x_2,\omega x_3).
\end{equation}

\section*{Appendix B: Quartic terms in the scalar potential}

Quartic terms in the scalar potential, which contain only Higgs triplets,
can be categorized
into three classes. To write some of the invariant terms, we define
the following quantities.
\begin{eqnarray}
&& (\xi\xi)\equiv\xi_4\xi_4+\xi_5\xi_5+\xi_6\xi_6,\quad
(\xi\xi)^\prime\equiv\xi_4\xi_4+\omega^2\xi_5\xi_5+\omega \xi_6\xi_6,
\nonumber \\ &&
(\xi\xi)^{\prime\prime}\equiv\xi_4\xi_4+\omega\xi_5\xi_5+\omega^2\xi_6\xi_6,
\quad
(\xi\xi^\dagger)\equiv\xi_4\xi_4^\dagger+\xi_5\xi_5^\dagger+\xi_6\xi_6^\dagger,
\nonumber \\ &&
(\xi\xi^\dagger)^{\prime}\equiv\xi_4\xi_4^\dagger
+\omega^2\xi_5\xi_5^\dagger+\omega\xi_6\xi_6^\dagger,\quad
(\xi\xi^\dagger)^{\prime\prime}\equiv\xi_4\xi_4^\dagger
+\omega\xi_5\xi_5^\dagger+\omega^2\xi_6\xi_6^\dagger,
\nonumber \\ &&
(\xi^\dagger\xi^\dagger)\equiv(\xi\xi)^\dagger,\quad
(\xi^\dagger\xi^\dagger)^\prime\equiv((\xi\xi)^{\prime\prime})^\dagger,\quad
(\xi^\dagger\xi^\dagger)^{\prime\prime}\equiv((\xi\xi)^{\prime})^\dagger.
\end{eqnarray}
Below we list all the distinct quartic terms in the scalar potential, which
are formed with only Higgs triplets of the MW model.
If a term is not self-adjoint, hermitian conjugate
of that should be included in the potential.
\begin{eqnarray}
&& [{\rm Tr}(\xi_1^\dagger\xi_1)]^2,~
[{\rm Tr}(\xi_2^\dagger\xi_2)]^2,~
[{\rm Tr}(\xi_3^\dagger\xi_3)]^2,~
[{\rm Tr}((\xi^\dagger\xi))]^2,~
{\rm Tr}(\xi_1^\dagger\xi_1){\rm Tr}(\xi_2^\dagger\xi_2),~
\nonumber \\ &&
{\rm Tr}(\xi_1^\dagger\xi_1){\rm Tr}(\xi_3^\dagger\xi_3),~
{\rm Tr}(\xi_1^\dagger\xi_1){\rm Tr}((\xi^\dagger\xi)),~
{\rm Tr}(\xi_2^\dagger\xi_2){\rm Tr}(\xi_3^\dagger\xi_3),~
{\rm Tr}(\xi_2^\dagger\xi_2){\rm Tr}((\xi^\dagger\xi)),~
\nonumber \\ &&
{\rm Tr}(\xi_3^\dagger\xi_3){\rm Tr}((\xi^\dagger\xi)),~
{\rm Tr}(\xi_1^\dagger\xi_2){\rm Tr}(\xi_1^\dagger\xi_3),~
{\rm Tr}(\xi_1^\dagger\xi_2){\rm Tr}(\xi^\dagger_2\xi_1),~
{\rm Tr}(\xi_1^\dagger\xi_2){\rm Tr}(\xi^\dagger_3\xi_2),~
\nonumber \\ &&
{\rm Tr}(\xi_1^\dagger\xi_2){\rm Tr}((\xi^\dagger\xi)^{\prime\prime}),~
{\rm Tr}(\xi_3^\dagger\xi_1){\rm Tr}(\xi^\dagger_1\xi_3),~
{\rm Tr}(\xi_3^\dagger\xi_1){\rm Tr}(\xi^\dagger_3\xi_2),~
{\rm Tr}(\xi_3^\dagger\xi_1){\rm Tr}((\xi^\dagger\xi)^{\prime\prime}),~
\nonumber \\ &&
{\rm Tr}(\xi_2^\dagger\xi_3){\rm Tr}(\xi^\dagger_3\xi_2),~
{\rm Tr}(\xi_2^\dagger\xi_3){\rm Tr}((\xi^\dagger\xi)^{\prime\prime}),~
{\rm Tr}((\xi^\dagger\xi)^\prime){\rm Tr}((\xi^\dagger\xi)^{\prime\prime}),
\nonumber \\ &&
{\rm Tr}(\xi_5^\dagger\xi_6){\rm Tr}(\xi^\dagger_6\xi_5)+
{\rm Tr}(\xi_6^\dagger\xi_4){\rm Tr}(\xi^\dagger_4\xi_6)+
{\rm Tr}(\xi_4^\dagger\xi_5){\rm Tr}(\xi^\dagger_5\xi_4),
\nonumber \\ &&
[{\rm Tr}(\xi_5^\dagger\xi_6)]^2+
[{\rm Tr}(\xi_6^\dagger\xi_4)]^2+
[{\rm Tr}(\xi_4^\dagger\xi_5)]^2.
\end{eqnarray}
\begin{eqnarray}
&&{\rm Tr}(\xi_1^\dagger\xi_1^\dagger){\rm Tr}(\xi_1\xi_1),~
{\rm Tr}(\xi_1^\dagger\xi_1^\dagger){\rm Tr}(\xi_2\xi_3),~
{\rm Tr}(\xi_1^\dagger\xi_1^\dagger){\rm Tr}((\xi\xi)),~
{\rm Tr}(\xi_2^\dagger\xi_3^\dagger){\rm Tr}(\xi_2\xi_3),~
\nonumber \\ &&
{\rm Tr}(\xi_2^\dagger\xi_3^\dagger){\rm Tr}((\xi\xi)),~
{\rm Tr}((\xi^\dagger\xi^\dagger)){\rm Tr}((\xi\xi)),~
{\rm Tr}(\xi_1^\dagger\xi_3^\dagger){\rm Tr}(\xi_1\xi_3),~
{\rm Tr}(\xi_1^\dagger\xi_3^\dagger){\rm Tr}(\xi_2\xi_2),~
\nonumber \\ &&
{\rm Tr}(\xi_1^\dagger\xi_3^\dagger){\rm Tr}((\xi\xi)^{\prime\prime}),~
{\rm Tr}(\xi_2^\dagger\xi_2^\dagger){\rm Tr}(\xi_2\xi_2),~
{\rm Tr}(\xi_2^\dagger\xi_2^\dagger){\rm Tr}((\xi\xi)^{\prime\prime}),~
{\rm Tr}((\xi^\dagger\xi^\dagger)^\prime){\rm Tr}((\xi\xi)^{\prime\prime}),~
\nonumber \\ &&
{\rm Tr}(\xi_1^\dagger\xi_2^\dagger){\rm Tr}(\xi_1\xi_2),~
{\rm Tr}(\xi_1^\dagger\xi_2^\dagger){\rm Tr}(\xi_3\xi_3),~
{\rm Tr}(\xi_1^\dagger\xi_2^\dagger){\rm Tr}((\xi\xi)^\prime),~
{\rm Tr}(\xi_3^\dagger\xi_3^\dagger){\rm Tr}(\xi_3\xi_3),~
\nonumber \\ &&
{\rm Tr}(\xi_3^\dagger\xi_3^\dagger){\rm Tr}((\xi\xi)^\prime),~
{\rm Tr}((\xi^\dagger\xi^\dagger)^{\prime\prime}){\rm Tr}((\xi\xi)^\prime),~
\nonumber \\ &&
{\rm Tr}(\xi_5^\dagger\xi_6^\dagger){\rm Tr}(\xi_5\xi_6)+
{\rm Tr}(\xi_6^\dagger\xi_4^\dagger){\rm Tr}(\xi_6\xi_4)+
{\rm Tr}(\xi_4^\dagger\xi_5^\dagger){\rm Tr}(\xi_4\xi_5).
\end{eqnarray}
\begin{eqnarray}
&&{\rm Tr}(\xi_1^\dagger\xi_1\xi_1\xi_1^\dagger),~
{\rm Tr}(\xi_1^\dagger\xi_1\xi_2^\dagger\xi_2),~
{\rm Tr}(\xi_1^\dagger\xi_1\xi_2\xi_2^\dagger),~
{\rm Tr}(\xi_1\xi_1^\dagger\xi_2^\dagger\xi_2),~
{\rm Tr}(\xi_1\xi_1^\dagger\xi_2\xi_2^\dagger),~
{\rm Tr}(\xi_1^\dagger\xi_1\xi_3^\dagger\xi_3),~
\nonumber \\ &&
{\rm Tr}(\xi_1^\dagger\xi_1\xi_3\xi_3^\dagger),~
{\rm Tr}(\xi_1\xi_1^\dagger\xi_3^\dagger\xi_3),~
{\rm Tr}(\xi_1\xi_1^\dagger\xi_3\xi_3^\dagger),~
{\rm Tr}(\xi_1^\dagger\xi_1(\xi^\dagger\xi)),~
{\rm Tr}(\xi_1^\dagger\xi_1(\xi\xi^\dagger)),~
{\rm Tr}(\xi_1\xi_1^\dagger(\xi^\dagger\xi)),~
\nonumber \\ &&
{\rm Tr}(\xi_1\xi_1^\dagger(\xi\xi^\dagger)),~
{\rm Tr}(\xi_2^\dagger\xi_2\xi_2\xi_2^\dagger),~
{\rm Tr}(\xi_2^\dagger\xi_2\xi_3^\dagger\xi_3),~
{\rm Tr}(\xi_2^\dagger\xi_2\xi_3\xi_3^\dagger),~
{\rm Tr}(\xi_2\xi_2^\dagger\xi_3^\dagger\xi_3),~
{\rm Tr}(\xi_2\xi_2^\dagger\xi_3\xi_3^\dagger),~
\nonumber \\ &&
{\rm Tr}(\xi_2^\dagger\xi_2(\xi^\dagger\xi)),~
{\rm Tr}(\xi_2^\dagger\xi_2(\xi\xi^\dagger)),~
{\rm Tr}(\xi_2\xi_2^\dagger(\xi^\dagger\xi)),~
{\rm Tr}(\xi_2\xi_2^\dagger(\xi\xi^\dagger)),~
{\rm Tr}(\xi_3^\dagger\xi_3\xi_3\xi_3^\dagger),~
{\rm Tr}(\xi_3^\dagger\xi_3(\xi^\dagger\xi)),~
\nonumber \\ &&
{\rm Tr}(\xi_3^\dagger\xi_3(\xi\xi^\dagger)),~
{\rm Tr}(\xi_3\xi_3^\dagger(\xi^\dagger\xi)),~
{\rm Tr}(\xi_3\xi_3^\dagger(\xi\xi^\dagger)),~
{\rm Tr}((\xi^\dagger\xi)(\xi\xi^\dagger)),~
{\rm Tr}((\xi^\dagger\xi)(\xi^\dagger\xi)),~
\nonumber \\ &&
{\rm Tr}(\xi_1^\dagger\xi_2\xi_1^\dagger\xi_3),~
{\rm Tr}(\xi_1^\dagger\xi_2\xi_3\xi_1^\dagger),~
{\rm Tr}(\xi_2\xi_1^\dagger\xi_1^\dagger\xi_3),~
{\rm Tr}(\xi_2\xi_1^\dagger\xi_3\xi_1^\dagger),~
{\rm Tr}(\xi_1^\dagger\xi_2\xi_1\xi_2^\dagger),~
{\rm Tr}(\xi_2\xi_1^\dagger\xi_2^\dagger\xi_1),~
\nonumber \\ &&
{\rm Tr}(\xi_1^\dagger\xi_2\xi_3^\dagger\xi_2),~
{\rm Tr}(\xi_1^\dagger\xi_2\xi_2\xi_3^\dagger),~
{\rm Tr}(\xi_2\xi_1^\dagger\xi_3^\dagger\xi_2),~
{\rm Tr}(\xi_2\xi_1^\dagger\xi_2\xi_3^\dagger),~
{\rm Tr}(\xi_1^\dagger\xi_2(\xi^\dagger\xi)^{\prime\prime}),~
{\rm Tr}(\xi_1^\dagger\xi_2(\xi\xi^\dagger)^{\prime\prime}),~
\nonumber \\ &&
{\rm Tr}(\xi_2\xi_1^\dagger(\xi^\dagger\xi)^{\prime\prime}),~
{\rm Tr}(\xi_2\xi_1^\dagger(\xi\xi^\dagger)^{\prime\prime}),~
{\rm Tr}(\xi_3^\dagger\xi_1\xi_3\xi_1^\dagger),~
{\rm Tr}(\xi_1\xi_3^\dagger\xi_1^\dagger\xi_3),~
{\rm Tr}(\xi_3^\dagger\xi_1\xi_3^\dagger\xi_2),~
{\rm Tr}(\xi_3^\dagger\xi_1\xi_2\xi_3^\dagger),~
\nonumber \\ &&
{\rm Tr}(\xi_1\xi_3^\dagger\xi_3^\dagger\xi_2),~
{\rm Tr}(\xi_1\xi_3^\dagger\xi_2\xi_3^\dagger),~
{\rm Tr}(\xi_3^\dagger\xi_1(\xi^\dagger\xi)^{\prime\prime}),~
{\rm Tr}(\xi_3^\dagger\xi_1(\xi\xi^\dagger)^{\prime\prime}),~
{\rm Tr}(\xi_1\xi_3^\dagger(\xi^\dagger\xi)^{\prime\prime}),~
\nonumber \\ &&
{\rm Tr}(\xi_1\xi_3^\dagger(\xi\xi^\dagger)^{\prime\prime}),~
{\rm Tr}(\xi_2^\dagger\xi_3\xi_2\xi_3^\dagger),~
{\rm Tr}(\xi_3\xi_2^\dagger\xi_3^\dagger\xi_2),~
{\rm Tr}(\xi_2^\dagger\xi_3(\xi^\dagger\xi)^{\prime\prime}),~
{\rm Tr}(\xi_2^\dagger\xi_3(\xi\xi^\dagger)^{\prime\prime}),~
\nonumber \\ &&
{\rm Tr}(\xi_3\xi_2^\dagger(\xi^\dagger\xi)^{\prime\prime}),~
{\rm Tr}(\xi_3\xi_2^\dagger(\xi\xi^\dagger)^{\prime\prime}),~
{\rm Tr}((\xi^\dagger\xi)^\prime(\xi^\dagger\xi)^{\prime\prime}),~
{\rm Tr}((\xi^\dagger\xi)^\prime(\xi\xi^\dagger)^{\prime\prime}),~
{\rm Tr}((\xi\xi^\dagger)^\prime(\xi^\dagger\xi)^{\prime\prime}),~
\nonumber \\ &&
{\rm Tr}((\xi\xi^\dagger)^\prime(\xi\xi^\dagger)^{\prime\prime}),~
{\rm Tr}(\xi_1\xi_1\xi_2^\dagger\xi_3^\dagger),~
{\rm Tr}(\xi_1\xi_1\xi_3^\dagger\xi_2^\dagger),~
{\rm Tr}(\xi_1\xi_1(\xi^\dagger\xi^\dagger)),~
{\rm Tr}(\xi_2\xi_3(\xi^\dagger\xi^\dagger)),~
\nonumber \\ &&
{\rm Tr}(\xi_3\xi_2(\xi^\dagger\xi^\dagger)),~
{\rm Tr}((\xi\xi)(\xi^\dagger\xi^\dagger)),~
{\rm Tr}(\xi_1\xi_2\xi_3^\dagger\xi_3^\dagger),~
{\rm Tr}(\xi_2\xi_1\xi_3^\dagger\xi_3^\dagger),~
{\rm Tr}(\xi_1\xi_2(\xi^\dagger\xi^\dagger)^{\prime\prime}),~
\nonumber \\ &&
{\rm Tr}(\xi_2\xi_1(\xi^\dagger\xi^\dagger)^{\prime\prime}),~
{\rm Tr}(\xi_3\xi_3(\xi^\dagger\xi^\dagger)^{\prime\prime}),~
{\rm Tr}((\xi\xi)^\prime(\xi^\dagger\xi^\dagger)^{\prime\prime}),~
{\rm Tr}[(\xi_5^\dagger\xi_6)^2+(\xi_6^\dagger\xi_4)^2
+(\xi_4^\dagger\xi_5)^2],~
\nonumber \\ &&
{\rm Tr}[\xi_5\xi_5^\dagger\xi_6\xi_6^\dagger
+\xi_6\xi_6^\dagger\xi_4\xi_4^\dagger+\xi_4\xi_4^\dagger\xi_5\xi_5^\dagger],~
{\rm Tr}[\xi_5^\dagger\xi_5\xi_6^\dagger\xi_6+
\xi_6^\dagger\xi_6\xi_4^\dagger\xi_4+\xi_4^\dagger\xi_4\xi_5^\dagger\xi_5],~
\nonumber \\ &&
{\rm Tr}[\xi_6\xi_5\xi_6^\dagger\xi_5^\dagger+
\xi_4\xi_6\xi_4^\dagger\xi_6^\dagger+\xi_5\xi_4\xi_5^\dagger\xi_4^\dagger],~
{\rm Tr}[\xi_5\xi_5\xi_6^\dagger\xi_6^\dagger+
\xi_6\xi_6\xi_4^\dagger\xi_4^\dagger+\xi_4\xi_4\xi_5^\dagger\xi_5^\dagger].
\end{eqnarray}

\end{document}